\documentclass[amsmath,amssymb,aps,prfluids]{revtex4-2}
\usepackage{graphicx}
\usepackage[colorlinks=true]{hyperref}
\usepackage{bm}
\usepackage[dvipsnames]{xcolor}

\newcommand{\change}[1]{{ #1}}

\begin{document}

\title{\change{Self-organisation of auto-phoretic suspensions in confined shear flows}}%

\author{Prathmesh Vinze}%
\affiliation{LadHyX, CNRS--Ecole Polytechnique, Institut Polytechnique de Paris, 91120 Palaiseau, France}
\author{Sebastien Michelin}%
\email[Electronic address: ]{sebastien.michelin@polytechnique.edu}
\affiliation{LadHyX, CNRS--Ecole Polytechnique, Institut Polytechnique de Paris, 91120 Palaiseau, France}
 
\date{\today}
\begin{abstract}
Janus phoretic particles exploit chemical energy stored in their environment to produce mechanical work on the surrounding fluid and  self-propel.  These  active  particles  modify  and  respond  to  their  hydrodynamic  and  chemical environments, thus providing them with a sensibility to external flows and other particles.  These chemical and hydrodynamic interparticle interactions are known to lead to non-trivial collective behaviour within such biological or synthetic active suspensions (e.g., cluster formation of phoretic particles or bacterial swarming). Recent experiments and analysis have demonstrated that  the response of active suspensions to shear flows is non-trivial and can, in fact, lead to significant reductions in viscosity due to the energy conversion at microscopic scales. In  this  work,  we  numerically  analyse using a continuum kinetic model the  dynamics and response to  shear  of   dilute  and  confined  suspensions  of chemotactic phoretic particles, that reorient and drift toward the chemical solutes released by their neighbors. We show that a 1D transient steady distribution driven by the effect of confinement is a common feature for the confinement and shear rate intensities considered. This 1D state is stable for strong confinement and thus observed in the long-term dynamics in sufficiently narrow channels. For wider channels, the transient state  becomes  unstable  to  streamwise  perturbations  due  to  the  chemotactic  instability,  leading  to  the formation of particle aggregates along the channel's walls. Their relative arrangements and dynamics are determined by the relative influence of shear intensity and chemotaxis and critically condition the suspension's dynamics and particle-induced flows. In a second step, the feedback effect on the flow and effective viscosity of the self-organised suspension are considered. We  show  that  the  induced  flow  and, consequently,  its  rheological  behaviour  strongly  depend  on  the  self-organisation  regime, and therefore on the interplay of confinement, shear and chemotaxis.
\end{abstract}

\maketitle

\section{Introduction}


Understanding the spontaneous self-organisation of large numbers of individually-powered and mobile agents, or \emph{active matter}, has recently emerged amongst the most active research fields in soft matter physics~\cite{marchetti2013,Ramaswamy2010}, at the interface with applied mathematics, fluid mechanics and biophysics because of its fundamental interest and potential applications~\cite{Gao2014,Carlsen2014,Nelson2010}. To self-propel, individual units convert energy (usually chemical) into mechanical work.  This definition of \emph{active} systems covers a broad range of characteristic lengthscales, from a few micrometres (bacterial suspensions~\cite{Pedley1992,mendelson1999}) to meters (e.g. animal herds~\cite{ginelli2015}, fish schools~\cite{partridge1982} or bird flocks~\cite{aplin2014}). 

Our focus here is on the former, where interacting agents are micron-sized and evolve in a suspending Newtonian fluid (i.e. ``wet active matter''). Such microswimmers can be broadly classified into two main categories: i) microorganisms (i.e. living systems such as bacteria, algae \change{and sperm cells}) and ii) synthetic microswimmers (i.e. engineered particles such as Janus colloids~\cite{Golestanian2007}, \emph{active} drops~\cite{Michelin2023} and \change{Quincke rollers~\cite{bricard2013}.}). At such  tiny scales, inertial effects are negligible, and swimmers' motion and interactions are dominated by viscous effects, introducing some well-known  constraints on the swimming strokes such as their non-reciprocity~\cite{purcell1977life}. To achieve this, living cells and organisms exploit irreversibility in the beating pattern of deformable flagella or cilia~\cite{lauga2009}. Reproducing in the lab such complex deformations and stroke patterns at microscopic scales is particularly challenging but possible using macroscopic forcing (e.g. magnetic field~\cite{Dreyfus2005,Babataheri2011}). 
Another important issue concerns the individual powering of these units: relying on a common directional macroscopic forcing (e.g. magnetic) indeed introduces a bias that tends to dominate inter-particle interactions, and interferes with the self-organisation. 

A popular alternative instead converts physico-chemical energy of the \emph{local} suspending fluid into mechanical work, e.g. via catalytic reactions of suspended solute species at the chemically-coated particles' surface~\cite{Howse2007,Moran2017}. Directional self-propulsion then relies on the phoretic drift of the particles~\cite{Anderson1989} in self-generated solute gradients produced by their front-back design asymmetry (Janus systems, \cite{Golestanian2007}). Despite the impressive diversity in their chemical nature and reactivity, most phoretic systems essentially rely on two main common properties of their surface: a physico-chemical \emph{activity}, namely the ability to produce, consume or alter a chemical solute, and a \emph{mobility} which enables them to convert local physico-chemical surface gradients into phoretic slip flows generated by the local differential interactions of solute and solvent molecules with the particles' surface~\cite{Anderson1989}.

Janus particles, as any other living or synthetic self-propelled swimmer, stir and perturb the surrounding flow, resulting in the hydrodynamic drift and reorientation of their neighbors~\cite{lauga2009}. These long-ranged hydrodynamic interactions amongst microswimmers are known to produce complex behaviour in large suspensions, including  pattern formation ~\cite{Pedley1992,Sokolov2009}, induced flows on lengthscales much larger than individual swimmers~\cite{mendelson1999,Dombrowski2004,dunkel2013}, active turbulence~\cite{alert2022} and enhanced particle diffusion~\cite{kim2004,leptos2009}.

Additionally, their chemical activity and mobility provide them with the ability to interact chemically over long ranges, through the gradients they create on their neighbors' environment: the particles thus not only swim under the effect of their own asymmetric properties, but also drift and reorient depending on the relative positioning of the surrounding particles. These chemical interactions open the route for biased motion of the swimmers under the influence of ``external'' chemical gradients, i.e. \emph{chemotaxis}~\cite{Adler1975}. Biological microswimmers exploit this ability to communicate and complete complex tasks such as targeting of inflammation/infectious sites by immune cells ~\cite{Petri2018}, locating mammalian/non-mammalian eggs for fertilization~\cite{eisenbach1999} and migrating towards a food source or away from a poisonous environment as a survival strategy~\cite{SOURJIK2012}. Biased motion for biological microswimmers is typically achieved via chemically-driven changes in their stochastic tumbling rate which increases or decreases depending on whether the cell senses deteriorating or improving conditions. Over long time scales, this results in a biased motion towards attractant-rich regions. Lacking bio-chemical sensors and complexity, Janus particles instead exploit asymmetric surface slip flows generated by their biased chemical environment to reorient along or against the local chemical gradient~\cite{Tatulea-Codrean2018,Kanso2019}. Thus, Janus particles interact via both chemical and hydrodynamical signatures~\cite{Sen2009} similar to some biological swimmers~\cite{Budrene1991,berg1972chemotaxis}.  
Modelling the two mechanisms (run/tumble and particle reorientation) gives rise to qualitatively similar results for chemotactic suspension behaviour on timescales much larger than the swimmers' tumbling rate~\cite{lushi2013}. When the solute mediating the interactions is directly produced or consumed by the particles, the suspension is termed \emph{auto-chemotactic}: particles not only follow or avoid their neighbors' proximity but also their own chemical footprint. A specific tendency of these suspensions is known as the formation of \emph{asters} as a result of a generic chemotactic instability~\cite{saha2014}, similar to their biological counterparts~\cite{Budrene1991}. 

A large part of the early work on active suspensions has focused on their spontaneous organisation in unbounded and quiescent flows, thus neglecting any potential environmental forcing or coupling, and focusing specifically on the intrinsic suspensions dynamics \cite{Saintillan2008,subramanian2009}. Yet, in order to achieve targeted applications (e.g. in biomedicine~\cite{Gao2014,Carlsen2014,Nelson2010}), control strategies must be obtained in realistic environments that feature confinement (i.e. the presence of bounding walls) and/or non-uniform background flows, reflecting an external mechanical actuation of the system.  Motivated by this observation, this work focuses on the response of a suspension of Janus particles under the dual influence of varying strength of confinement and externally-applied shear. Response to shear further provides some insight into the effective rheology of the suspension~\cite{Rafai2010,Lopez2015,Burkholder2020}. Focusing specifically on Janus phoretic swimmers, Ref.~\cite{traverso_michelin_2022} considered the dynamic response and effective rheology of a dilute suspension in pressure-driven Poiseuille flow between two rigid walls, identifying five different regimes depending on the relative strength of the flow-inducing pressure drop.

Microswimmers are able to interact with confining boundaries through their stirring of the surrounding fluid. The constraint of a no-slip boundary condition at finite distance introduces an additional flow perturbation that modifies their dynamics and results in non-trivial behaviour, even at an individual scale. Among other effects, previous studies report  attraction and reorientation towards a wall~\cite{spagnolie2012}, scattering of biflagellate microswimmers from circular surfaces~\cite{Contino2015,Lushi2017}, microswimmer trapping using a stationary obstacle~\cite{spagnolie2015}, or the observation of different states depending on catalyst coverage for Janus particles~\cite{ibrahim2016,choudhary2021} and wall accumulation at suspension scale~\cite{rothschild1963}. 

Wall accumulation of swimmers is well documented~\cite{rothschild1963,berke2008,Li2011}: accumulating spermatozoa at the rigid walls~\cite{rothschild1963} play an important role in mammalian reproduction~\cite{suarez2006}. Interestingly, in contrast with short-term dynamics of individual swimmers, at time scales much larger than the typical reorientation or run-and-tumble rate of the swimmers,  hydrodynamics is not even necessary to explain such wall accumulation, which can then be seen as a result of the coupling of self-propulsion and confinement~\cite{costanzo2012,ezhilan_saintillan_2015}. The characteristic length-scale of the confining boundaries further plays an important role in controlling the collective dynamics of microswimmer suspensions~\cite{Lushi2014,Wioland2013}: in the experimental results of Wioland.\textit{et.al}~\cite{wioland2016}, the distribution of bacteria suspended in a fluid transitions from a complex 2D structure to a streamwise independent 1D distribution as the confinement strength is increased. \change{This points to the critical role of the confinement intensity, whose effect on the suspension's self-organisation  is a central motivation for the present work.}

The presence of a background flow also results in a distinct swimmer behavior, such as the directional bias of \emph{E.coli} in pressure-driven flows~\cite{figueroa2015}. Such rheotaxis can be understood from the dual effect of steric interactions and background shear, which aligns the bacteria  against the flow due to their elongated shape~\cite{kaya2012}. Similar results were also reported for sperm cells in shear flows~\cite{omori2016}. Synthetic swimmers such as Janus phoretic particles also display  rheotactic behavior, which was  reported for spherical and rod-shaped catalytic colloids in confined shear flows~\cite{Uspal2015,Brosseau2019}. More recently, Traverso and Michelin~\cite{traverso_michelin_2022} showed that Janus particles also polarize against the flow, like \emph{E.coli}, under strong background Poiseuille flow. For spherical particles, such rheotaxis can be understood from the competition of wall normal polarization and background vorticity. For elongated particles, for which steric effects play an important role, Ref.~\cite{Brosseau2019} showed that rheotaxis can also be tuned using the detailed activity and actuation patterns of the active particles.

Understanding the behaviour and response of chemotactic suspensions in shear flows is critical for their applications as fluids of non-trivial or even tunable effective viscosity~\cite{Lopez2015,martinez2020}. 
It is now well established that elongated pusher-type swimmers tend to decrease the effective viscosity of the suspension~\cite{Lopez2015,gachelin2014}, while spherical puller-type swimmers (e.g. algae) tend to increase it~\cite{Rafai2010}\change{, thanks to the microscopic stresses they exert on the surrounding fluid and the shear alignment of the swimmers}. 
The rheological response is qualitatively captured by theoretical models proposed by Saintillan~\cite{Saintillan2010} and Hatwalne \emph{et al.}~\cite{Hatwalne2004} \change{ for unconfined suspensions. These models are furthermore not applicable to a suspension of spherical Janus particles which do not experience  geometric shear alignment.} The rheology of such suspensions in Poiseuille flows was \change{therefore} recently analysed by Ref.~\cite{traverso_michelin_2022}, that reported a net reduction in the global effective viscosity \change{at low shear rates}, consistently with the observations on other microswimmer suspensions~\cite{gachelin2014}.


\change{Two main types of models  have been used to analyse the collective dynamics and macroscopic response of microswimmer suspensions, namely particle-based and kinetic models. The former describe and track each individual particle and are thus particularly well-suited to describe dense suspensions since they can resolve detailed inter-particle modeling. Particle-based  dynamic simulations reproduce at least qualitatively the experimental behaviour of such suspensions~\cite{saintillan2012emergence,rojas2021} and are therefore invaluable since they can provide scaling laws by statistical analysis.} However, such simulations are also computationally expensive, especially for large system sizes \change{making it more difficult to perform systematic analyses of different physical effects}. In contrast,  kinetic models consider directly the global evolution of the suspension via a probability density distribution of finding particles with specific orientation at a given point~\cite{subramanian2009,Saintillan2008}. \change{Accurate description of particle couplings in dense systems becomes however more complex, and in practice, }kinetic models are mostly used in the dilute limit, when typical inter-particle distances are sufficiently large~\cite{Saintillan2008,Saintillan2015}. \change{In this approach, which is the one chosen in this work, hydro-chemical coupling of the particles is accounted for }through the influence on individual particles of hydrochemical mean fields, forced by the individual hydrodynamic and chemical footprints of the particles~\cite{liebchen2015,Traverso2020}.

\change{The primary goal of this work is to understand the dual effect of background shear flow and confinement. While previous studies have explored the effect of shear on self-organisation of active suspensions, most of them focused on pressure-driven flows~\cite{ezhilan_saintillan_2015,traverso_michelin_2022} with non-uniform shear profile. As a result, the regions where background shear is most significant overlaps with the regions where the wall influence dominates, leading to complex dynamics~\cite{traverso_michelin_2022} and making it harder to decipher the role of each effect. To avoid this, we focus in this work on a simple shear profile (Couette), such that the effect of shear is similar in the entire domain, whether close or far from the boundaries. In contrast, the effect of confinement is profoundly non-uniform across the channel, affecting mostly the regions away from the centerline. Additionally, in this configuration, the asymmetric shear flow results in a differential advection of the particles on both sides of the channel and in horizontal interactions between the aggregates which has not been reported earlier. So far existing studies  have either explored fixed confinement effect~\cite{ezhilan_saintillan_2015,traverso_michelin_2022}, or varying confinement for non-chemotactic suspensions~\cite{theillard2017}; by varying the confinement strength, we obtain here a better insight into the role of confinement on the suspension dynamics with respect to the intrinsic chemotactic behaviour and report that the confinement strength stabilizes the 1D regime observed at short times, due to strong transverse gradients. Finally, the flow induced by the particles' self-organisation and the suspension's rheology is discussed and a minimalistic model is proposed.}

The manuscript is organized as follows: Sec.~\ref{sec:formulation} describes the physical model and summarizes the governing equations, characteristic scalings and numerical approach for their resolution. The self-organisation behaviour is then analysed in detail in Sec.~\ref{sec:dist_dyn}, where three different long-term regimes are identified based on the two control parameters of the problem, namely the strength of the uniform background shear and the degree of confinement. Based on this understanding, Sec.~\ref{sec:sus_rheo} then proposes an overview of the resulting effective viscosity of the suspension, focusing specifically on the different flow patterns induced by the particle distribution and forcing in the different dynamical regimes reported in Sec.~\ref{sec:dist_dyn}. This is followed by an analysis of the suspension rheology in Sec.~\ref{sec:sus_rheo}, which discusses the different induced flows observed corresponding to different regimes. Sec.~\ref{sec:conc} finally summarizes the main conclusions of our work and presents some future perspectives.

\section{Modeling dilute suspensions of sheared phoretic particles}
\label{sec:formulation}

\begin{figure}
  \includegraphics{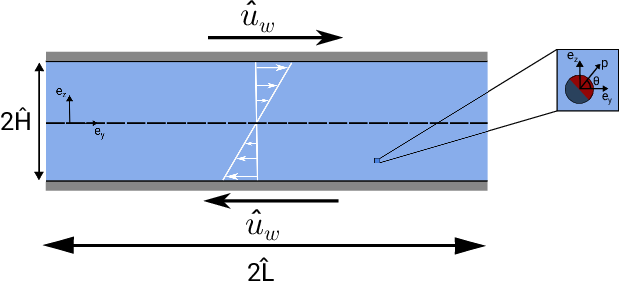}
  \caption{Problem Schematic. A dilute suspension of Janus particle placed between two flat walls. The walls move opposite to each other to establish a simple shear flow}
\label{fig:Schematic}  
  \end{figure}

\subsection{Physical model}
\subsubsection{Description}
This work considers the self-organisation and response to shear of a suspension of self-propelling Janus particles of radius $\hat{R}$ in a Newtonian fluid (viscosity $\hat\eta$) confined between two flat walls separated by a distance $2\hat{H}$;  both walls move in opposite directions at a speed of $\hat{u}_w$ : in the absence of any particle, this would establish  a steady Couette shear flow with uniform shear rate $\hat{\gamma}=\hat{u}_w/\hat{H}$ as illustrated in  Fig.~\ref{fig:Schematic}, a quantity referred to in the following as background shear rate. All dimensional quantities are represented with a `hat( $\hat{}$ )' in order to distinguish  them later on from their dimensionless counterparts. Particles interact with a chemical solute of concentration $\hat{C}(\hat{\mathbf{x}},\hat{t})$ through two fundamental physico-chemical properties: a chemical \emph{activity} $\hat{A}(\hat{\mathbf{x}})$ (namely the ability to produce or consume solute) and a mobility $\hat{M}(\hat{\mathbf{x}})$ that converts surface tangential gradients of solute into slip flows along their boundaries, the combination of which enables the particle to set up local chemical gradients through which it can self-propel~\cite{Golestanian2007,Moran2017}. The particles considered in this work consist of two chemically-homogeneous hemispheres, and $\bm{p}$ is thus the particle orientation while $\hat{A}^\pm$ and $\hat{M}^\pm$ denote the sum $(+)$ and front-back difference $(-)$ in activity and mobility of the two hemispheres.

\subsubsection{Kinetic model of dilute suspensions}
When the suspension is sufficiently dilute \change{(i.e. when $\hat{R}$ is much smaller than $\hat{H}$ and the typical interparticle distance)}, a classical approach to model its dynamics is based on the probability density function $\hat{\Psi}(\bm{\hat{x},p},\hat{t})$ of finding a particle at position $\bm{\hat{x}}$ with orientation $\bm{p}$ defined by the axis of symmetry as shown in Fig.~\ref{fig:Schematic} at time $\hat{t}$~\cite{Saintillan2010,subramanian2009}. 

In general, the probability density $\hat{\Psi}$ is a function of six independent variables, namely the three spatial coordinates, two angular coordinates and time. Finding the suspension dynamics in $3D$, is therefore computationally intensive; in order to gain some physical insight on the suspension dynamics while keeping computational costs manageable, we restrict ourselves to the analysis of the two-dimensional problem, where $\hat{\Psi}$ only depends on two spatial and a single angular coordinate ($\hat{y},\hat{z},\theta$) (Fig.~\ref{fig:Schematic}) and time $\hat{t}$. Such two dimensional reduction has been made in previous studies and showed qualitatively accurate with respect to experiments \cite{Saintillan2008,Lushi2012,Lushi2014}. 

The local particle density $\hat\Phi(\bm{x},t)$  and polarisation $\bm{n}(\bm{\hat{x}},\hat{t})$ are defined as, 
\begin{equation}
    \hat\Phi=\int_{\Omega}\hat{\Psi}(\hat{\bm{x}},\bm{p},\hat{t})d\bm{p},\qquad \bm{n}=\frac{1}{\hat{\Phi}}\int_{\Omega}\hat{\Psi}(\hat{\bm{x}},\bm{p},\hat{t})\bm{p}d\bm{p}
\label{eqn:Phi}
\end{equation}
where  $\Omega$ spans all possible orientations (unit circle in $2D$); the mean particle density in the suspension  is then given by 
\begin{equation}
    \hat N=\frac{1}{\hat S}\int_{\hat{\mathcal{S}}}\hat{\Phi}d\hat{S}=\frac{1}{\hat S}\int_{\hat{\mathcal{S}}} \int_{\Omega}\hat{\Psi}(\bm{\hat{x},p},\hat{t})d\bm{p}d\hat{S}.
\end{equation}
The conservation of particles writes locally as a Smoluchowski equation for $\hat{\Psi}$\cite{Saintillan2013}, 
\begin{equation}
    \frac{\partial\hat{\Psi}}{\partial t}=-\hat{\nabla}_x \cdot (\dot{\bm{\hat{x}}}\hat{\Psi})-\hat{\nabla}_{p}\cdot(\dot{\bm{p}}\hat{\Psi}) 
    \label{eqn:psi}
\end{equation}
with $\hat{\nabla}_x$ and $\hat{\nabla}_p$, the differential operators in space and orientation. 

The translation and rotation fluxes on the right-hand side of Eq.~\eqref{eqn:psi} are obtained from the translation $\bm{\dot{\hat{x}}}$ and rotation  velocity $\dot{\bm{p}}$  of an isolated particle placed in the hydrodynamic and chemical mean-fields $\hat{\bm{u}}(\bm{\hat{x}})$ and $\hat{C}(\bm{\hat{x}})$, corrected for the diffusion contribution~\cite{traverso_michelin_2022}
\begin{equation}
    \dot{\bm{\hat{x}}}=\hat{U}_0\bm{p}+\hat{\bm{u}}+\hat{\chi}_t \hat{\nabla}_x\hat{C}-\hat{D}_x\nabla_x[\text{ln}(\hat{\Psi})],
\end{equation}

\begin{equation}
    \dot{\bm{p}}=\frac{1}{2}\hat{\omega}\times\bm{p}+\hat{\chi}_r(\bm{p}\times \hat{\nabla}_x \hat{C})\times \bm{p}-\hat{D}_p\hat{\nabla}_p[\text{ln}(\hat{\Psi})].
\end{equation}
Here, $\hat{U}_0$ is the self-propulsion velocity of the Janus particle,  $\hat\chi_t$ and $\hat\chi_r$ its translation and rotation phoretic mobilities, $\hat{\omega}=\hat{\nabla}_x\times \bm{\hat{u}}$ the local vorticity, and $\hat{D_x},\hat{D_p}$ are translation and orientation particle diffusivities.

The specific values of $\hat{U}_0,\hat{\chi}_t,\hat{\chi}_r$ depend on detailed physico-chemical properties (surface activity and surface mobility) and coating patterns. For the hemispherical particles considered here, $\hat{U}_0,\hat{\chi}_t,\hat{\chi}_r$ are given by~\cite{Traverso2020,traverso_michelin_2022} 
\begin{equation}
\hat{U}_0=\frac{\hat{A}^{-}\hat{M}^{+}}{8\hat{D}_c},\; \hat{\chi}_t=-\frac{\hat{M}^{+}}{2},\; \hat{\chi}_r=\frac{9\hat{M}^{-}}{16\hat{R}}\cdot
\end{equation}    
 The suspension is bounded at the top and bottom by impermeable walls, so that the wall-normal component of the probability flux $\hat{\Psi}\dot{\hat{\bm{x}}}$ must vanish there
\begin{equation}
\bigg(\hat{U}_0\sin\theta+\chi_t\frac{\partial \hat{C}}{\partial \hat{z}} \bigg)\hat{\Psi}=\hat{D}_x\frac{\partial \hat{\Psi}}{\partial \hat{z}}\qquad \textrm{at   }\hat{z}=\pm\hat{H}. 
\end{equation}

\subsubsection{Hydrodynamic problem}
Finding $\hat{\Psi}$ also requires solving for the chemical and hydrodynamic fields $\hat{C}$ and $\hat{\bm{u}}$. The  small lengthscales of typical experiments ($\hat{H}\sim 10^{-4} - 10^{-3} m$)\cite{Lopez2015,Wioland2013,Gachelin2013} guarantee that inertial effects on the flow are negligible ($\mbox{Re}\sim O(10^{-2})$). The flow velocity $\bm{\hat{u}(x)}$ and pressure $\hat{q}(\bm{x})$ therefore satisfy Stokes equations  forced by the moving boundaries and the individual hydrodynamic active stresses $\bm{\hat{S}}(\bm{\hat{x}},\hat{t})$ exerted by the different particles:
\begin{equation}
\hat{\nabla}_x \cdot \bm{\hat{u}}=0,\qquad
-\hat{\eta} \hat{\nabla}^2_x \bm{\hat{u}} + \hat{\nabla}_x \hat{q}=\hat{\nabla}_x\cdot\bm{\hat{S}}.
\end{equation}     
The no-slip boundary condition at the walls imposes 
\begin{equation}
\bm{\hat{u}}= \pm \hat{u}_w\bm{e}_y \;\; \text{at} \;\; \hat{z}=\pm \hat{H}.
\end{equation}
Here, $\bm{\hat{S}}$ is the local stress induced by the Janus particles which is evaluated by taking orientation average of the stresslet produced by a single swimmer oriented along $\bm{p}$ as\cite{traverso_michelin_2022,Saintillan2008}
\begin{equation}
\bm{\hat{S}}(\bm{\hat{x}},\hat{t})=\int_\Omega\bm{\hat{S}_p}(\bm{\hat{x},p},\hat{t})\hat{\Psi} d\bm{p}.
\end{equation} 
The stresslet $\bm{\hat{S}_p}$ produced by the slip forcing at the surface of a single Janus particle consists of two parts: (i) the response $\bm{\hat{S}_s}$ to self-induced chemical gradients (i.e. the particle's own activity)  and (ii) the response $\bm{\hat{S}_e}$ to chemical gradients induced  at the particle position by its surroundings and neighbors. The strength of both parts depend on physio-chemical properties and are given for the present two-dimensional problem by~\cite{traverso_michelin_2022,Traverso2020}, 
%
\begin{equation}
\bm{\hat{S}_s}= \hat{\alpha}_s \bigg(\bm{pp-\frac{I}{2}} \bigg)\quad\textrm{with  }\hat{\alpha_s}=-\frac{10\pi\kappa\hat{\eta} \hat{R}^2\hat{M}^{-}\hat{A}^{-}}{\hat{D}_c}
\label{eqn:stress_self}
\end{equation} 

\begin{equation}
\bm{\hat{S}_e}= \hat{\alpha}_e \left[\bm{\hat{G}p+p\hat{G}}+\bm{(\hat{G}\cdot p)}\left(\bm{pp}-\frac{3\bm{I}}{2}\right) \right] \quad\textrm{with  } \hat{\alpha_e}=\frac{15}{8}\pi\hat{\eta} \hat{R}^2 \hat{M}^{-}
\label{eqn:stress_ex}
\end{equation}
with $\kappa \approx 0.0872$ and $\hat{\bm{G}}$ the local external solute gradient. 


\subsubsection{Chemical problem}
Each particle's activity perturbs the solute field, and at the scale of a single particle, the disturbance field is classically obtained in terms of spherical harmonics~\cite{Golestanian2007}. However, at the scale of the whole suspension, only the long range term (slowest decaying term) survives and corresponds to the net consumption/production by the particle with a rate $2\pi\hat{R}^2 \hat{A}^+$. In the following, and without any loss of generality, the considered particles are net solute producers ($\hat{A}^{+}>0$), and the solute concentration $\hat{C}(\bm{\hat{x}},\hat{t})$ satisfies an advection diffusion equation forced by the particles' individual solute production and a  relaxation towards a background equilibrium, namely 
\begin{equation}
\frac{\partial \hat{C}}{\partial \hat{t}}+\bm{\hat{u}}\cdot \hat{\nabla}_x \hat{C}=\hat{D}_c\hat{\nabla}^2_x\hat{C}-\hat{\beta}\hat{C}+2\pi \hat{R}^2 \hat{A}^{+} \Phi
\end{equation}
The walls are chemically inert: $\hat{C}$ must also satisfy a no-flux boundary condition, 
\begin{equation}
\frac{\partial \hat{C}}{\partial \hat{z}}=0 \; \text{at} \; \hat{z}=\pm \hat{H}
\label{eqn:noflux_sol}
\end{equation}

\subsection{Dimensionless equations}
In the following, the channel half-width ($\hat{H}$) and corresponding solute diffusion time ($\hat{H}^2/\hat{D}_c$) are chosen as characteristic length- and timescales, and  $\hat{D}_c/\hat{H}$ and $\hat{\eta} \hat{D}_c/\hat{H}^2$ are the corresponding  velocity and pressure scales, respectively. The characteristic concentration scale $\hat{C}_{ch}= \hat{H} \hat{A}^{+}/\hat{D}_c\zeta$ is obtained by balancing the solute production ($\hat{N} \hat{R}^2\hat{A}^{+}$) with solute diffusion ($\hat{D}_c \hat{C}_{ch}/\hat{H}^2$) where $\zeta=(\hat{H}\hat{N}\hat{R}^2)^{-1}$ is the ratio of the characteristic suspension scale $(\hat{N}\hat{R}^2)^{-1}$~\cite{Saintillan2008} to the channel half-width $\hat{H}$. The parameter $\zeta$ is a relative measure of the suspension's length scale to the channel width and thus defines the degree of confinement. 
The probability density is normalized by the mean number density $\hat{N}$. 
Equation~\eqref{eqn:psi} remains unchanged with dimensionless translation and rotation fluxes now given by 
\begin{equation}
    \bm{\dot{x}}=u_0 \bm{p}+ \bm{u} + \frac{\xi_t}{\zeta}\nabla_x C-d_x\nabla_x[\text{ln}\Psi],
\end{equation}

\begin{equation}
    \dot{\bm{p}}=\frac{1}{2}\omega\times\bm{p}+\frac{\xi_r}{\rho\zeta}(\bm{p}\times \nabla_x C)\times \bm{p}-d_p\nabla_p[\text{ln}(\Psi)]
\end{equation}
where $\rho=\hat{R}/\hat{H}$ is the relative particle size, and $d_x=\hat{D}_x/\hat{D}_c$ and $d_p=\hat{D}_p\hat{H}^2/\hat{D}_c$, the ratios of particle translation or rotation diffusion to solute diffusion, respectively. Meanwhile, the dimensionless self-propulsion speed $u_0$, phoretic drift coefficient $\xi_t$ and chemotactic reorientation coefficient $\xi_r$ are 
\begin{equation}
u_0=\frac{\hat{A}^{-}\hat{M}^{+}\hat{H}}{8\hat{D}^2_c },\; \xi_t=-\frac{\hat{M}^{+}\hat{A}^{+}\hat{H}}{2\hat{D}^2_{c}},\; \xi_r=\frac{9\hat{M}^{-}\hat{A}^{+}\hat{H}}{16\hat{D}^2_c}
\end{equation} 
Using  Eq.~\eqref{eqn:noflux_sol}, the boundary condition for $\Psi$ can be written as 
\begin{equation}
u_0 \sin\theta\Psi=d_x\frac{\partial \Psi}{\partial z}. 
\label{eqn:no_flux_psi}
\end{equation}
Stokes equation are written in non-dimensional form as
\begin{equation}
\nabla_x\cdot \bm{u}=0,\qquad 
- \nabla^2_x \bm{u} + \nabla_x q=\nabla_x\cdot\bm{S}
\label{eqn:Stokes}
\end{equation}
with  boundary conditions 
\begin{equation}
\bm{u}= \pm \gamma \bm{e_z} \;\; \text{at} \;\; z=\pm 1,
\end{equation}
and the strength of the dimensionless stresslets are now given by
\begin{equation}
\alpha_s= \frac{640\pi\kappa}{9}\frac{\xi_ru_0}{\xi_t\zeta} \hspace{0.01\textwidth} \text{, } \alpha_e=\frac{10\pi}{3}\frac{\xi_r}{\zeta^2}\cdot \label{eq:adim_stresslet}
\end{equation} 
Note here that the dimensionless ratio $\gamma / u_0$ determines the relative strength of the externally-imposed shear flow to the particles' self-propulsion: for $\gamma \ll u_0$ (weak shear), the background forcing is not sufficiently strong to influence significantly the particle transport; in contrast, for strong shear forcing  ($\gamma \gg u_0$), the particles behave as if they were passive.

Finally, the advection-diffusion equation becomes in dimensionless form
\begin{equation}
\frac{\partial C}{\partial t}+\bm{u}\cdot \nabla_x C=\nabla^2_x C-\beta C+2\pi \Phi
\label{eqn:adv_diff}
\end{equation}
where $\beta^{-1/2} =\hat{l}_c/\hat{H}$ with $\hat{l}_c=\sqrt{\hat{D}_c}/\hat\beta$ the  dimensionless screening length introduced by the relaxation of the concentration (essentially the range of chemical influence of individual particles). The no flux boundary condition for the solute remains unchanged in dimensionless form. 

\subsection{Numerical simulations}
\label{sec:Num_scheme}
The numerical framework of Ref.~\cite{traverso_michelin_2022} is adapted here to the present work's configuration and forcing. Following this work, Eqs.~\eqref{eqn:psi}, \eqref{eqn:Stokes} and \eqref{eqn:adv_diff} are solved numerically using a pseudo-spectral scheme with Chebyshev discretization along the vertical direction ($z$, with $N_z$ number of modes) and Fourier decomposition in the periodic horizontal ($y$, with $N_y$ number of modes) and angular directions ($\theta$, with $N_{\theta}$ number of modes). A resolution of $N_y=N_z=64$ and $N_{\theta}=32$ is chosen for the results presented here, for which a relative error of $10^{-3}$ on the effective viscosity (i.e. integrated force on the top plate) is measured with respect to a refined discretisation with $N_y=N_z=128$ and $N_{\theta}=64$. \change{ For $\zeta=1$, the computational box chosen for the results reported here has dimensions $L_y=2\hat{L}/\hat{H}=6\pi$ in the streamwise direction and $L_z=2$ in the vertical direction. We observe that doubling $L_y$ does not change the self-organisation results reported in Sec.\ref{sec:dist_dyn} which suggests that the box size does not affect the system dynamics, at least provided $L_z\ll L_y$.}


Simulations are initiated starting from a nearly uniform and isotropic state given by $\Psi(\bm{x,p},t=0) =\frac{1}{2\pi}+\epsilon \tilde{\Psi}(\bm{x,p}) $ with $\epsilon\ll1$, and the corresponding purely diffusive solution for ${C}$ is used initially. \change{The full governing equations are then marched in time by treating the diffusive terms semi-implicitly and the non-linear terms explicitly so that we solve a set of 1D Helhmholtz equation at each time step.} The non-linear boundary condition, Eq.~\eqref{eqn:no_flux_psi} couples all Chebyshev modes in $z$ for each $(y,\theta)$-mode, and is treated numerically by transforming it formally into a Neumann boundary condition on $\Psi$. The solution is then iterated until its convergence (defined when the norm of the relative error is less than $10^{-4}$
) at each time step (typically up to 6 iterations). The Chebyshev \textit{tau}-method decouples the odd modes with the even modes for the set of $N_yN_{\theta}/4$ 1D Helmholtz equations that are solved at each iteration and time step, thereby greatly reducing the computational cost.
Lastly, at each time step, the Stokes equations, Eq.\eqref{eqn:Stokes} are solved using the \textit{influence-matrix} method which ensures mass conservation locally to machine precision~\cite{Kleiser1980}.   

\subsection{Selection of the physical parameter values}
\begin{table}
\begin{ruledtabular}
\begin{tabular}{ccc}
  \textrm{Symbol}&
\textrm{Physical Parameter }&
\textrm{Magnitude estimate}\\
\colrule
$\hat{R}$ & Particle Radius~\cite{Howse2007,Sen2009}  & $10^{-6}m$\\
$\hat{H}$  & Channel Width~\cite{Lopez2015} & $10^{-4}m$ \\
$\hat{U_0}$  & Swimming Speed~\cite{Paxton2004} & $10^{-6}ms^{-1}$\\ 
$\hat{D}_p$   & Rotational Diffusion coefficient~\cite{valiev} & $10^{-1}s^{-1}$\\
$\hat{D}_c$   & Solute Diffusion coefficient~\cite{subczynski1984diff} & $10^{-9}m^2s^{-1}$\\
$\hat{D}_x$   & Particle Diffusion coefficient~\cite{Howse2007} & $10^{-11}m^2s^{-1}$\\
\end{tabular}
\end{ruledtabular}
\caption{\label{tab:table0}%
\change{Dimensional parameters of the system together with their typical order of magnitude in experiments and references where such an estimate can be drawn from.} }
\end{table}
The problem is described by several different non-dimensional parameters relating the properties of the system and particles\change{, and in the following, we specifically focus in the following on the role of confinement $\zeta$ and shear $\gamma$. To estimate these, dimensional parameter values are chosen so as }to be relevant to existing experiments \change{(see Table.~\ref{tab:table0})}. Janus particles are typically micron-sized~\cite{Howse2007,Sen2009} ($\hat{R}\sim 10^{-6}$m) and  swim at a speed of a few bodylengths/sec~\cite{Paxton2004} ($\hat{U}_0\sim 10^{-6}$m/s). Typical microfluidic channels feature sub-millimeter widths ($\hat{H}\sim 10^{-4} - 10^{-3}$ m)\cite{Lopez2015,Gachelin2013} and the solute diffusion coefficient for small molecules such as dissolved oxygen gas is $\hat{D}_c\sim 10^{-9}$ m$^2$/s~\cite{subczynski1984diff}. As a result, the non-dimensional swimming velocity of the particles $u_0\sim O(1)$, therefore setting $u_0=0.5$ in the simulation ensures physical relevancy. \change{We consider chemotactic particles (i.e. that reorient along local chemical gradients), and fix $\xi_r/\rho=1.25$ and $\xi_t=-0.5$; we note that such particles leave a pusher-like hydrodynamic footprint on the surrounding fluid, Eq.~\eqref{eq:adim_stresslet}. With these values, the effects of phoretic drift, chemotactic reorientation and self-propulsion} are of similar magnitude resulting in complex dynamics. A stronger self-propulsion velocity would prevent the particles to form aggregates  while a lower $u_0$ would delay the aggregates' formation~\cite{Traverso2020}. The rotational diffusion coefficient for the particles can be estimated based on temperature ($\hat{T}$), radius ($\hat{R}$) and viscosity ($\hat{\eta}$) by Einstein's relation as $\hat{D}_p=\hat{k}_{B}\hat{T}/(8\pi\hat{\eta}\hat{R}^3)\sim 10^{-1}$m$^2$s$^{-1}$ \cite{valiev} which results in dimensionless diffusion coefficient as $d_p=\hat{D}_p\hat{H}^2/\hat{D}_c=0.25$. Similarly, the effective translational diffusion coefficient for the particles can be estimated as $\hat{D}_x=\hat{k}_B \hat{T}/(8\pi\hat{\eta}\hat{R})+\hat{U}^2_0\hat{D}^{-1}_p/2\sim 10^{-11} $m$^2$s$^{-1}$~\cite{Howse2007}, so that $d_x=D_x/D_c=0.025$. Finally, setting $\beta=\pi/2$ results in a $O(1)$ dimensionless screening length thus ensuring that particles interact throughout the channel.

\section{Self-organisation dynamics}
\label{sec:dist_dyn}



\change{The suspension's self-organisation results from  different intrinsic effects (the particles' self-propulsion, their phoretic attraction/repulsion and chemotactic reorientation) and their competition with shear-induced rotation. In addition, the particles evolve in a confined setting and are transported by the flow. To shed a better light on the results presented in the rest of the paper, and understand how the dynamics of the present system arise from their interaction, we first describe how chemotaxis, flow forcing and confinement independently act on the suspension's organisation.}


Self-organisation of unbounded phoretic suspensions in quiescent flows is dominated by autochemotaxis, i.e the particles' ability to sense, reorient and migrate toward or away from a specific chemical signal, here generated by the chemical signatures of their neighbours.
 Such chemically-driven interactions are also known to play an important role in the self organisation of biological suspensions \cite{Adler1975,Petri2018}, where the microswimmers typically change their tumbling rate depending on a specific chemical cue to create an orientation bias towards the chemical source \cite{Budrene1991}. Such chemotactic self-organisation results in a variety of complex behaviour such as pattern formation~\cite{Budrene1991}, swarms~\cite{Jacob2004,ariel2013}, bacterial turbulence~\cite{dunkel2013}, etc.  
 Janus phoretic swimmers instead exploit a front-back asymmetric coating and the resulting polarity in their interaction with suspended solutes, in order to reorient along chemical gradients~\cite{Tatulea-Codrean2018,vinze2021}. Particle aggregation and cluster formation may result from such chemotactic interactions. 
 For net solute producers (${\hat{A}^{+}>0}$, as in the present configuration), any infinitesimal inhomogeneity in the spatial distribution of phoretic colloids  triggers more solute production and local solute accumulation in specific regions~\cite{Traverso2020,liebchen2015}.
The associated long-ranged chemical gradients generate an orientation bias towards those regions among the particles nearby, which cause their own swimming and accumulation in the regions of already-higher particle concentration (for positively-chemotactic particles). More solute is then generated there which results in a positive feedback loop and extension of the process throughout the domain~\cite{Traverso2020}. 

The characteristic timescale for such chemotactic clustering is $\tau_c\sim (\tau_\chi/\tau_\beta)\tau_s^\lambda$~\cite{Traverso2020}, where $\tau^\lambda_s\sim 1/(kU_0)$ is the characteristic time scale of self-propulsion over a perturbation wavelength $\lambda=2\pi /k$, $\tau_\chi\sim (kC_\textrm{ref}\chi_r)^{-1}$ is the typical scale for the chemical reorientation in the concentration gradient associated with the perturbation's spatial inhomogeneity, and $\tau_\beta\sim 1/\beta$ the characteristic relaxation time of the solute concentration in the bulk. Here $C_\textrm{ref}=HA^{+}/\zeta D_c$ is the characteristic concentration scale obtained by balancing solute production of the particles and the diffusive flux. The definition of $\tau_c$ with respect to the three time scales can be physically understood as follows: chemical reorientation polarises the suspension towards regions of excess solute toward which a majority of the (polar) particles swim. Clustering of the self-propelled particles take a time $\tau_s$ for fully-polarized particles. Polarisation, i.e. chemical reorientation however takes a finite amount of time and $\tau_\chi/\tau_\beta$ can be seen as a measure of how much polarized the suspension is able to get before the concentration perturbation triggering the reorientation  relaxes under the effect of chemical decay. This instability eventually saturates when the chemotactic flux is balanced by other processes such as diffusion and any potential repulsive phoretic drift within the chemical gradient, thus leading to the formation of high-density particle aggregates.  

The suspension dynamics and self organisation are also influenced by the presence of background (i.e. externally-imposed) flows; these not only advect particles differentially in non-uniform flows, but flow gradients also introduce a local reorientation/rotation of the particles (Faxen Laws).
For the anti-symmetric background flow imposed here (a simple shear flow), particles in the top half are advected in the opposite direction with respect to particles present in the bottom half, and the vorticity (and the clockwise induced rotation) is uniform throughout the channel (see Sec.~\ref{sec:Longterm}).

Lastly, the presence of walls impermeable to both particles and chemical solutes, result in their confinement and accumulation near the boundaries, a well-known feature of any (biological or synthetic) suspension of microswimmers~\cite{berke2008,li2009,Wioland2013,Gachelin2013}.


\subsection{Overview of the suspension dynamics}
\label{sec:SOdyn}
\label{sec:Overview}
\begin{figure}
\includegraphics[width=0.8\textwidth]{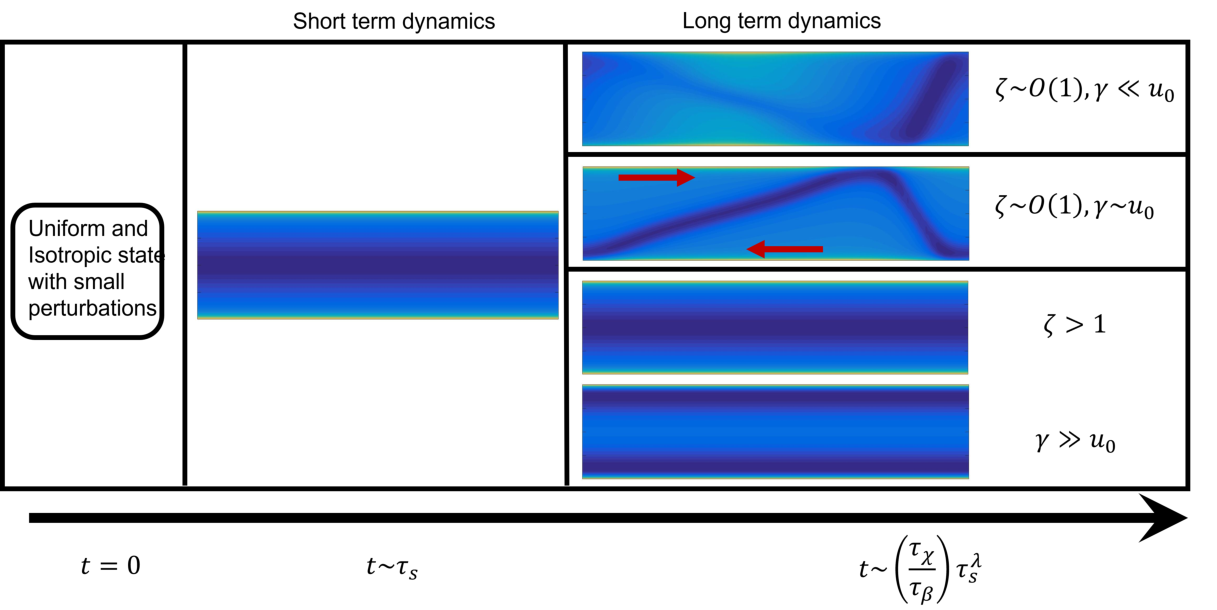}
\caption{Figure depicting the overview of self-organisation of the current system. Identification of three different regimes based on shear rate and confinement. Strong shear or strong confinement tends to stabilise the 1D state.}
\label{fig:overview}
\end{figure}

Starting from an initial perturbation of the isotropic initial condition, the  dynamics of the suspension can be decomposed into two successive phases occuring over two different and well-separated timescales: a short time scale associated with self-propulsion across the channel width, and a long time scale associated with self-organisation due to chemotactic instability (Fig.~\ref{fig:overview}). Its main features are outlined here before being discussed in more details in the next subsections.

At short times, as a result of the particles' self-propulsion and of their lateral confinement, the suspension quickly develops a transient 1D state (i.e. spatially invariant in the streamwise direction). This phase and the transient state it converges to, do not depend on the shear rate or confinement ratio  for the range of confinement and shear rates explored in this work(Fig.~\ref{fig:overview}). However, the clustering timescale suggests that the clustering time reduces with \change{decreasing} confinement. As a result, in the limit of unconfined suspensions, $\zeta \to 0$, the aggregation timescale becomes \change{shortest} resulting in clustering in the short-term itself. 

On the other hand, the long term solution results from the competition of chemotaxis, imposed shear and confinement; as such the long term dynamics \change{and three regimes are obtained}, as shown in table \ref{tab:table1}, depending on \change{ the intensity of confinement and shear}. These regimes can be distinguished by their  spatial distribution (1D or 2D) and temporal nature (steady/unsteady). \change{The ``steady'' nature of the converged state was checked for such situations by a doubling of the simulation time to ensure proper convergence.}

\begin{table}
\begin{ruledtabular}
\begin{tabular}{cccc}
  \multicolumn{2}{c}{\textrm{Control parameters}}& \multicolumn{2}{c}{\textrm{Long term response}}\\
  \colrule
  \textrm{Background Shear}&
\textrm{Confinement}&
\multicolumn{1}{c}{\textrm{Temporal}}&
\textrm{Spatial}\\
\colrule
Strong/Weak & Strong & Steady & 1D\\
Weak & Weak & Steady & 2D\\
Strong & Weak & Unsteady & 2D\\
\end{tabular}
\end{ruledtabular}
\caption{\label{tab:table1}%
Distinctive characteristics of the long-term dynamics of the chemotactic suspension for different level of confinement and background shear forcing. 
}
\end{table}

\subsection{Short term dynamics}
\label{sec:short_term}
\begin{figure}
\includegraphics[width=0.9\textwidth]{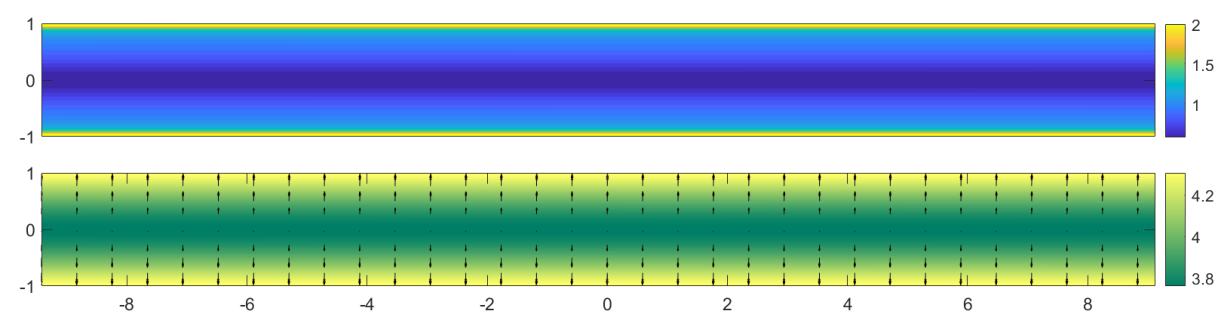}
\caption{Short-term particle (top) and solute distribution (bottom) obtained for $\zeta=1,\gamma=0.25$. The black arrows(bottom)  represent the local polarisation direction and magnitude of the particles.}
\label{fig:1Ddis}
\end{figure}

\begin{figure}
  \includegraphics[width=\textwidth]{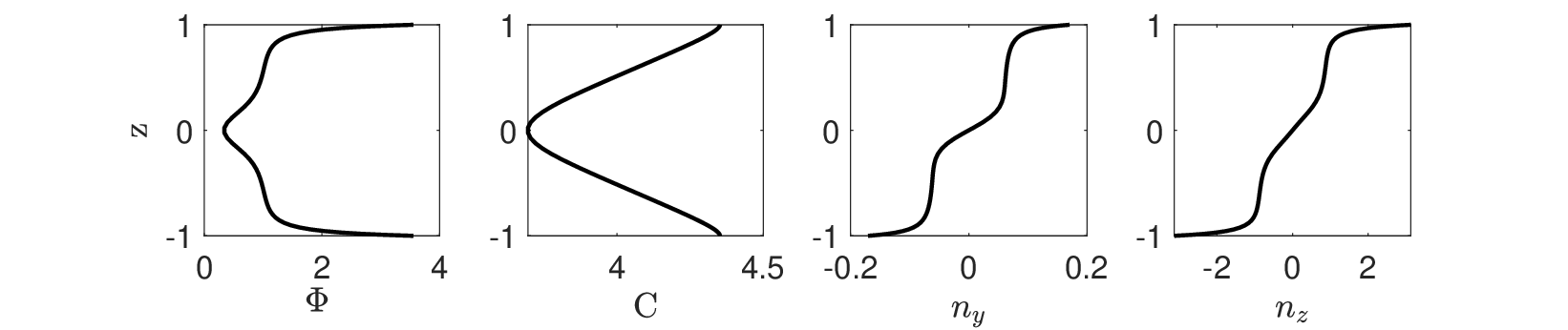}
\caption{ 1D profiles of the solute and particle concentrations, and  streamwise and wall-normal components of the particle polarisation in the transient regime. The 1D profiles are for $\zeta=1,\gamma=0.25$.}
\label{fig:1Ddis2}
\end{figure}
 Starting from the initial nearly-uniform and isotropic distribution, the suspension quickly relaxes to the 1D transient state shown in Fig.~\ref{fig:1Ddis}. Out of the three different effects driving the self-organisation of the suspension, namely the external shear, the chemotactic instability and self-propulsion across the channel width, the latter is associated with the shortest time scale $\tau_s^H\sim H/U_0$, and thus drives the dynamics of this early phase.

This 1D state is characterised by high particle densities near the boundaries (Fig.\ref{fig:1Ddis2}), a \change{rather intuitive} behaviour that is also well-established for suspensions of biological microswimmers~\cite{berke2008,li2009}: particles located in the vicinity of an impermeable boundary and oriented towards it are trapped there as they can only escape thanks to translational and rotational diffusion; instead particles oriented away from the boundary quickly swim away from this region.
This results in a strong wall polarisation $\bm{n}$,  Eq.~\eqref{eqn:psi}, within a thin boundary layer of particles near the channel walls (Fig.\ref{fig:1Ddis2}), whose thickness is proportional to rotational and translational diffusion and inversely proportional to self propulsion velocity~\cite{ezhilan_saintillan_2015}.

The particles \change{are net solute producers} ($A^+>0$), and their accumulation near the wall leads \change{to a locally-increased solute production near the impermeable walls, resulting in an accumulation of solute in the walls' vicinity}. Consequently, the solute distribution across the channel width is characterised by a V-shaped profile (Fig.\ref{fig:1Ddis2}) associated with strong chemical gradients pointing toward the boundaries.
This results in the formation of a strong solute gradient toward the boundary, which  polarises the suspension toward the nearest wall under the effect of chemical reorientation ($\xi_r>0$ here).
This reorientation combined with self-propulsion reinforces the particles' polarisation, accumulation and trapping near the boundary.  \change{The chemotactic behavior of the particles and their response to rapid spatial changes in the local solute gradient direction results in a divergent chemotactic flux and an additional local dip in the particle concentration profile near the channel center line, as seen on Fig.~\ref{fig:1Ddis2}. }

\subsection{Long term dynamics}
\label{sec:Longterm}
\begin{figure}
\includegraphics[width=0.75\textwidth]{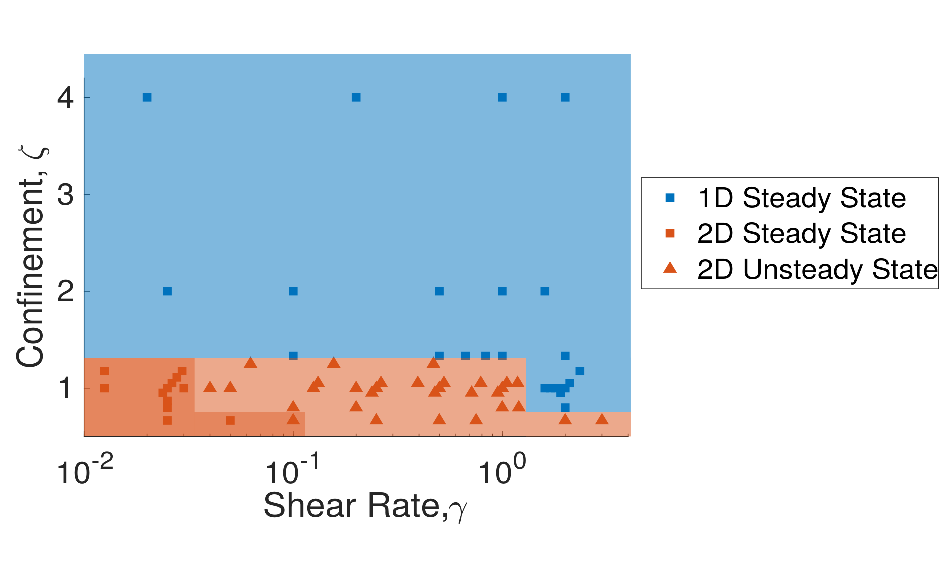}
\caption{Long-term dynamics of the chemotactic suspension for varying relative shear rate \change{$\gamma$} and degree of confinement \change{$\zeta=(\hat{H}\hat{N}\hat{R}^2)^{-1}$}. Colours indicate the nature of the particle distribution: 1D (blue) and 2D (orange). Different symbols are used for steady (square) and unsteady (triangle) regimes. }
\label{fig:phase}
\end{figure}

Depending on the background shear and confinement levels, the transient 1D state described in the previous section may be unstable with respect to (slower) streamwise perturbations under the effect of chemotactic clustering. In that case, the evolution of the system toward its long-term dynamics is driven by the chemotactic instability and the typical duration of this evolution thus scales as $\tau_c\sim \chi_r C_\textrm{ref}/U_0\beta$ (see section~\ref{sec:Overview}). The relevant characteristic scale for solute concentration remains the one used for non-dimensionalisation $C_\textrm{ref}=HA^{+}/\zeta D_c$.

The long term dynamics broadly divides into two different types of regimes, depending on the confinement level as summarized on the phase map of Fig.~\ref{fig:phase}.
For strong confinement (small channel width, $\zeta<1$), the behaviour of the system \change{remains that observed in the transient dynamics, namely the confinement-induced particle accumulation  near the wall . In that case,} the particle and solute distributions are independent of both \change{ $y$ and $t$ (steady 1D regime)}.

In contrast, when \change{the channel width} is large enough (i.e. low confinement, $\zeta>1$), the long term solution is characterised by the formation of aggregates along each wall, breaking the $y$-invariance of the solution, as a result of the chemotactic instability. In such regimes, the influence of the walls is weaker and the dynamics is thus more prominently driven by the intrinsic behaviour of the chemotactic suspension, \change{as for} unconfined suspensions~\cite{Lushi2012,Saintillan2013}. The particle and solute distributions are now \change{fully two-dimensional and can be steady or unsteady depending on the shear rate forcing}. 


These different regimes are presented and discussed in more detail in the following.
\subsubsection{Weak confinement}
\label{sec:Longterm_2D}
\begin{figure}
\includegraphics[width=0.9\textwidth]{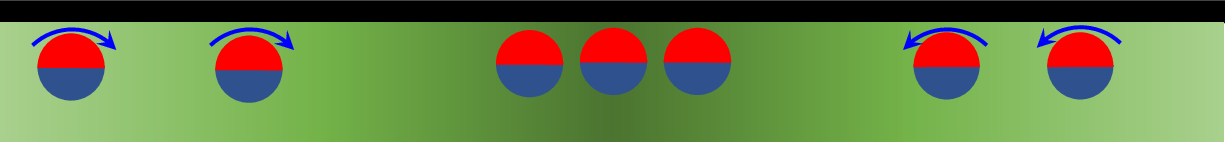}
\caption{1D boundary layer chemotactic destabilisation: a small disturbance in the particle distribution along the wall and the particles' activity introduce a local increase in solute concentration (green colour) and a small horizontal bias of the orientation of neighboring particles that start swimming toward and accumulating in this solute-rich region. }
\label{fig:pert}
\end{figure}
\begin{figure}
\includegraphics[width=0.9 \textwidth]{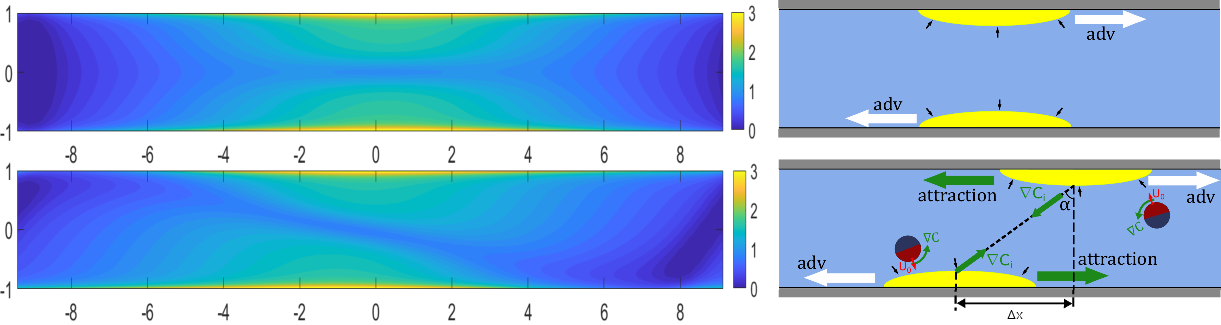}

\caption{(Left) Long term particle distribution for $\zeta=1$ and in the absence of external flow (top) and for ${\gamma=0.025}$ (bottom) 
. (Right) Schematic representation of the opposite-walls aggregates in the absence of background flow (top) and for weak shear flow (bottom). The black arrows represent the local polarisation of the particles, white arrows show the direction of background advection and green arrow indicate the direction of chemotactic bias (reorientation).}
\label{fig:2DSS}
\end{figure}

For weak confinement (or in the absence of any confinement, $\zeta<1$), the 1D transient state observed at short time becomes unstable with respect to two-dimensional (i.e. $y$-dependent) perturbations and evolves into the formation of 2D particles' aggregates. For confined suspensions, the 1D state is characterised by much higher density of particles near the walls and a strong polarisation of these particles towards the wall, as discussed in the previous section. This has two important consequences. First, the chemotactic instability and clustering develop preferentially within and along this concentration boundary layer. Additionnally, given the strong vertical chemical gradient of the 1D state, small disturbances in the solute concentration only significantly impact the horizontal concentration gradient and particle polarisation. These two effects result in the 1D version of the more general chemotactic instability reported for unconfined suspensions~\cite{Traverso2020} as shown schematically in Fig.~\ref{fig:pert}.

\change{In the absence of external flows, the particles form aggregates at regularly-spaced positions along each wall, determined by the dominant wavelength of the chemotactic instability. Inside each aggregate, particles remain mostly oriented toward the wall with a slight horizontal tilt to wards the center of the aggregate they belong to.} 
\change{For sufficiently wide channels (weak confinement), aggregates }along each wall only weakly influence each other, \change{yet, they introduce} a small bias in the horizontal solute gradient seen by an aggregate located close to the opposite wall.

This cross-channel chemotactic influence is unable to overcome the strong wall polarisation and strong vertical solute gradients seen by the particles, \change{but any offset of the aggregagtes on opposite walls introduces} a (very weak) horizontal bias  given by $|\nabla_x C_i| \sin \alpha$ where $\alpha$ is the relative position angle of the particle aggregates (Fig.~\ref{fig:2DSS}).
The horizontal bias coupled with self propulsion results in horizontal particle migration (Fig.\ref{fig:2DSS}, bottom right) along the walls, until the aggregates are placed symmetrically. This arrangement is an equilibrium position of the system in the long term in the absence of external flows.\\
 
\begin{figure}
  \includegraphics[width=\textwidth]{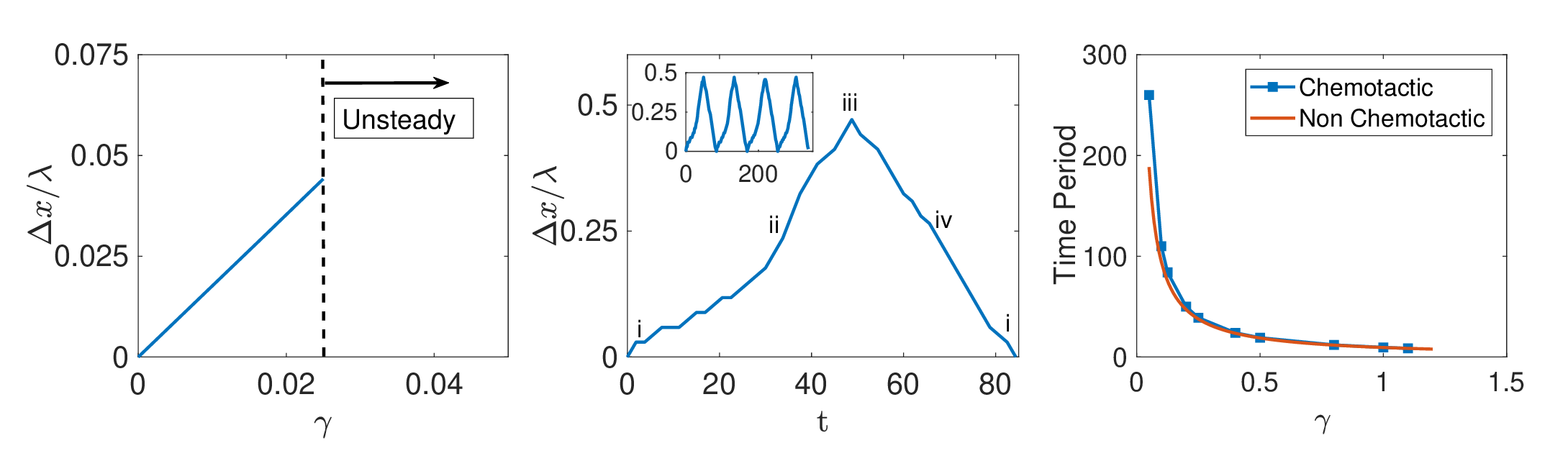}
\caption{(Left) Evolution of the steady minimum offset between aggregates on opposite walls($\Delta/\lambda x$) with the shear intensity(with $\lambda$ representing the wavelength of the most unstable mode), in the limit of weak shear and $\zeta=1$. (Centre) Time evolution of $\Delta x(t)/\lambda$ for the unsteady regime at $\zeta=1$ and sufficiently large shear forcing($\gamma=0.125$) with $\lambda$ representing the wavelength of the most unstable mode. (Right) Evolution with the shear rate intensity of the time period of the oscillations in the relative positioning of aggregates on the opposite walls for $\zeta=1$.}
\label{fig:Sep}
\end{figure}

Externally-imposed flows (and their gradient) transport both particles and solute in the streamwise direction.
For a symmetric flow such as Poiseullie flow, the particle aggregates on each wall are advected in the same direction resulting in a travelling wave solution at long times~\cite{traverso_michelin_2022}. For the present Couette flow configuration (homogeneous external shear), the aggregates on either walls are transported in opposite directions, and this anti-symmetric transport competes with the chemotactic clustering described above.
For relatively-weak flow forcing ($\gamma\ll u_0$), chemotaxis is strong enough to maintain a steady offset equillibrium of the aggregates (Fig.~\ref{fig:2DSS}): as the background forcing is weaker than self propulsion, the particles' positions remain trapped until diffusion enables them to escape.
Consequently, \change{aggregates are slightly offset horizontally, Fig.~\ref{fig:2DSS}, with an horizontal offset $\Delta x$ increasing} linearly with the shear intensity  (Fig.~\ref{fig:Sep})\change{: }the perturbation of the solute gradient magnitude seen by a given aggregate due to the \change{counterpart on the opposite wall is negligible}, and $\sin\alpha \approx \frac{\Delta x}{2H}$, resulting in a linear relationship between the chemotactic attraction and $\Delta x$, and thus with the convective transport by the background shear (Fig.~\ref{fig:2DSS}).

\begin{figure}
  \includegraphics[width=\textwidth]{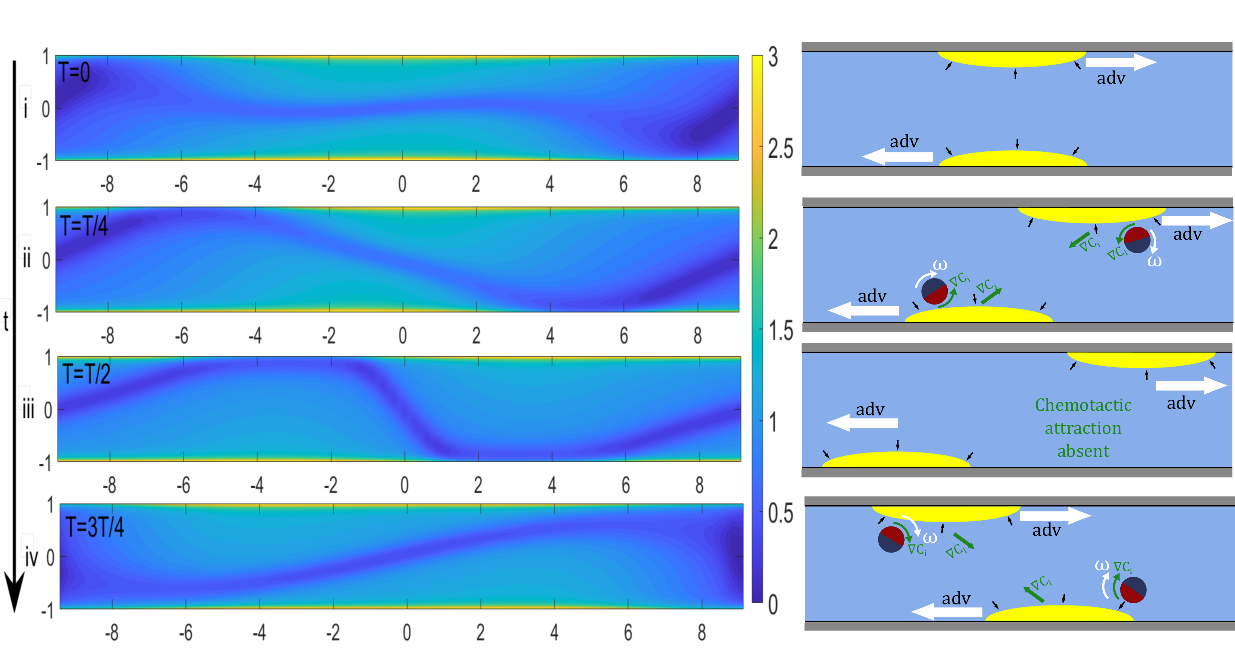}
  \caption{(Left) Evolution of the particle density in time over a period of the relative motion of aggregates on opposite walls for $\gamma=0.125$ and $\zeta=1$. (Right) Corresponding schematic representation of the position of the particles' aggregates. }
\label{fig:2D_evol}
\end{figure}

When the shear rate becomes large enough, the horizontal offset of the aggregates becomes significant and the magnitude of the perturbed concentration gradient($\nabla_x C_i$) responsible of their attraction decays as $O\big(\frac{1}{d^2}\big)$ (\change{aggregates are net solute sources}) where $d=\Delta x/\sin\alpha$ is the total distance between the two aggregates.
Beyond a critical horizontal separation (Fig.~\ref{fig:Sep}), the chemotactic attractive effect is not sufficient to balance the convective forcing, resulting in a continuous relative transport of the aggregates by the flow in the streamwise direction, and an unsteady but periodic dynamics (Fig.~\ref{fig:2D_evol}).\\

This periodic regime is however characterised by an asymmetric evolution of $\Delta x$ over one period (Fig.~\ref{fig:Sep}, centre), which can be understood by considering the relative direction of the chemotactic and convective forcings seen by the different moving aggregates, over a given period starting when aggregates from opposite walls are at their minimum distance ($\Delta x\approx 0$, Fig.~\ref{fig:2D_evol}i). During the first \change{half-period}, cross-channel chemotactic effects compete with particle and solute transport by the shear flow until they are perfectly offset from each other (i.e. maximum $\Delta x$, Fig.~\ref{fig:2D_evol}iii): this results, at least at first, in a slower relative motion of the aggregates in comparison with a purely convective transport. In contrast,  during the second half of the cycle, chemotactic attraction \change{by the closest opposite-wall aggregate} takes some time to build up (due to the large distance of the walls) and can not significantly enhance the transport velocity, even though it is acting now in the same direction as the convective forcing. This suggests, that once the chemotactic attraction is reversed (Figure~\ref{fig:2D_evol}ii), the aggregates are merely advected in opposite directions by the background shear flow, and the time taken to complete this second half of the period is identical to that for two non-chemotactic aggregates.
Overall, the total period of the oscillation is greater than for non chemotactic aggregates as the chemotactic attraction resists advection in the first half reducing the separation velocity (Fig.~\ref{fig:Sep}). This increase is most significant for low shear rate, as expected as chemotactic coupling is able to act longer (Fig.~\ref{fig:Sep}, right).

\subsubsection{Strong confinement}
\label{sec:Longterm1D}
\begin{figure}
\includegraphics[width=0.45\textwidth]{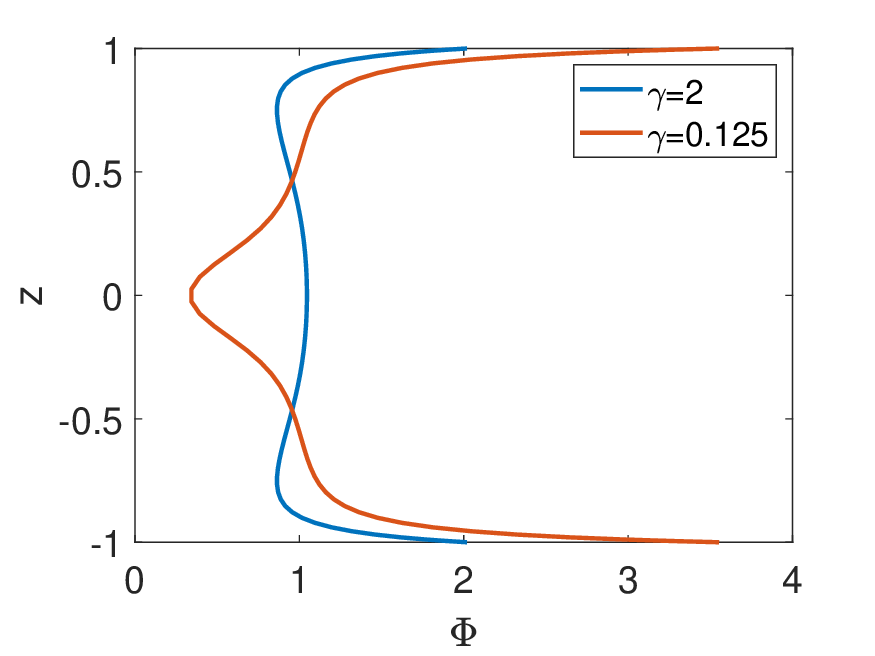}
\includegraphics[width=0.45\textwidth]{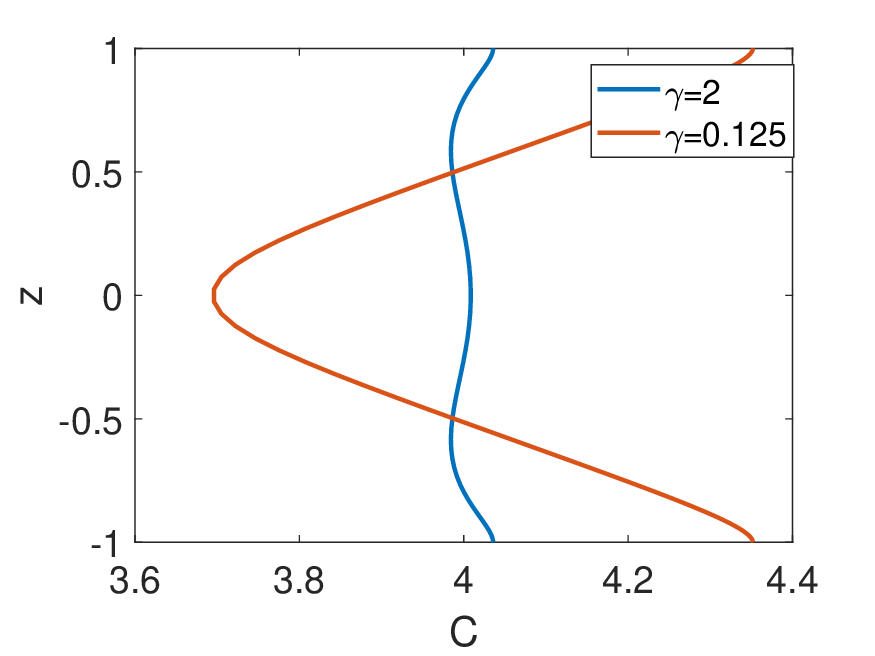}
\caption{Particle distribution (left) and solute concentration (right) observed in the strongly confined regime ($\zeta=2$) for low and high shear rates. }
\label{fig:1D2SS}
\end{figure}
  
\change{For stronger confinement, i.e. when the channel width is comparable to or smaller than the characteristic wave length of the chemotactic instability, the influence of the confining boundary  on the distribution of particles and solutes tends} to suppress the onset of the chemotactic instability of the 1D state observed at short times (Sec.~\ref{sec:short_term}).

\change{This was already} reported on other microswimmer systems in previous experimental~\cite{wioland2016,bricard2013} and numerical studies~\cite{theillard2017,theillard2019} -- and even in macroscopic systems \cite{Buhl2006}. Its main origin is the relative weakening of  horizontal gradients of solute in comparison with the strong vertical gradients. Consequently, the particles maintain a strong vertical polarisation and horizontal reorientation and polarisation is more difficult and unlikely\change{:  the net relative horizontal  displacement of the particles is then negligible. } As a result, \change{for strong confinement, the long term dynamics is invariant in $y$ (1D) and $t$ (steady), as in the short term}. The exact particle distribution of this 1D state depends on the specific value of shear rate considered, and may be significantly different from the transient regime characteristics; yet, all long-term regimes share common features regardless of the shear rate intensity, including the strong polarisation and accumulation of both solute and particles near the wall. The main characteristics of these 1D regimes are illustrated on Fig.~\ref{fig:1D2SS} for weak and strong shear, respectively. 


For low shear rates, the long term solution is in fact essentially identical to the transient solution, being characterised by high particle density at the walls and a strong reduction of the particle density near the channel centerline (see Sec.~\ref{sec:short_term}). This should be no surprise: for weak shear, the flow forcing is negligible and the dominant mechanisms leading to the self-organisation of the suspension are those intervening in the short-term (namely the impermeable wall boundaries), since the 2D chemotactic instability and clustering are suppressed. 

In contrast, for strong shear rates, background vorticity dominates the orientation dynamics resulting in the tumbling of particles (the particles are spherical) in the bulk of the channel where wall polarisation effects are weaker. This reduces or prevents the reorientation of particles present in the bulk in the direction of the closest wall under chemotactic effects, and maintains a larger concentration of the particles in the channel's bulk~ (Fig.~\ref{fig:1D2SS}). As a direct consequence, and due to the presence of this increased number of solute-producing particles, the chemical concentration is also higher further reducing the influence of chemotaxis toward the channel walls, flattening the solute distribution profile at high shear rates in comparison to the V-shaped profile observed at low shear values.
\subsection{\change{Linear Stability of the 1D equilibrium: a minimal model}}

In order to gain insights into the emergence of the different regimes described above as a result of the competition of confinement and shear with the chemotactic instability, we focus in this section on a minimalistic model based on a moment expansion of the probability density function. This model  includes qualitatively the main physical features of the problem and we show that it is able to capture at least qualitatively some of the complex suspension dynamics, such as  the long-term state for various shear rate and degree of confinement.
It follows in that regard an approach already used in existing works on  active suspensions~\cite{traverso_michelin_2022,ezhilan_saintillan_2015}.

Its central idea  is to reduce \change{the description of the probability distribution to its first orientational moments~\cite{Saintillan2013}, namely the particle density $\Phi$(zeroth moment) and its local orientation $\bm{n}$(first moment)}.
We refer the readers to appendix~\ref{sec:appendixA} for more details on the governing equation and its derivation. Such models are often applied to non-chemotactic elongated swimmers, whose  interactions depend critically on the second moment of $\Psi$~\cite{Saintillan2008,Saintillan2013}. In contrast, interactions among spherical Janus particles and with the background flow can already be included in a model involving only the zeroth (concentration) and first moments (orientation).

We noted that a full simulation of the system systematically predicted the emergence (at least transiently) of a 1D $y$-invariant state for all the simulations performed for the range of degree of confinement($\zeta$) and shear rate($\gamma$) considered in this work. Depending on the relative importance of confinement, shear and chemotaxis, this invariance along the horizontal was either maintained at long times or evolved toward steady/unsteady 2D regimes. We  interpret this long term evolution of the system as the result of the stability/instability of the system's steady 1D solution of the problem. Formulation of the regime selection as a simple eigenvalue problem is one of the main goal of the present reduced model for which the stability analysis can be carried out more easily.

\begin{figure}
    \includegraphics[width=1\textwidth]{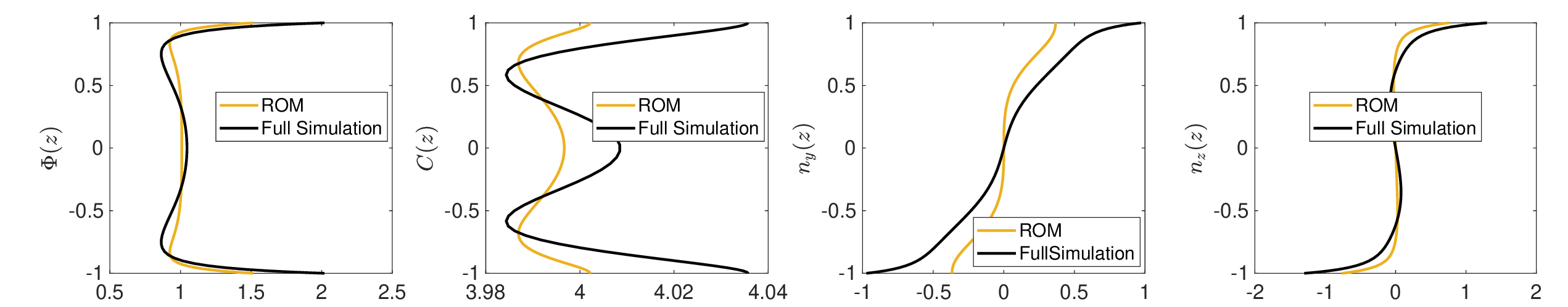}
    \caption{Comparisson between the 1D steady states obtained by solving 1D version of the reduced order equations to the full simulation for $\zeta=1$ and $\gamma=1.5$.}
    \label{fig:1D_SS_ROM}
\end{figure}

\begin{figure}
    \includegraphics[width=0.9\textwidth]{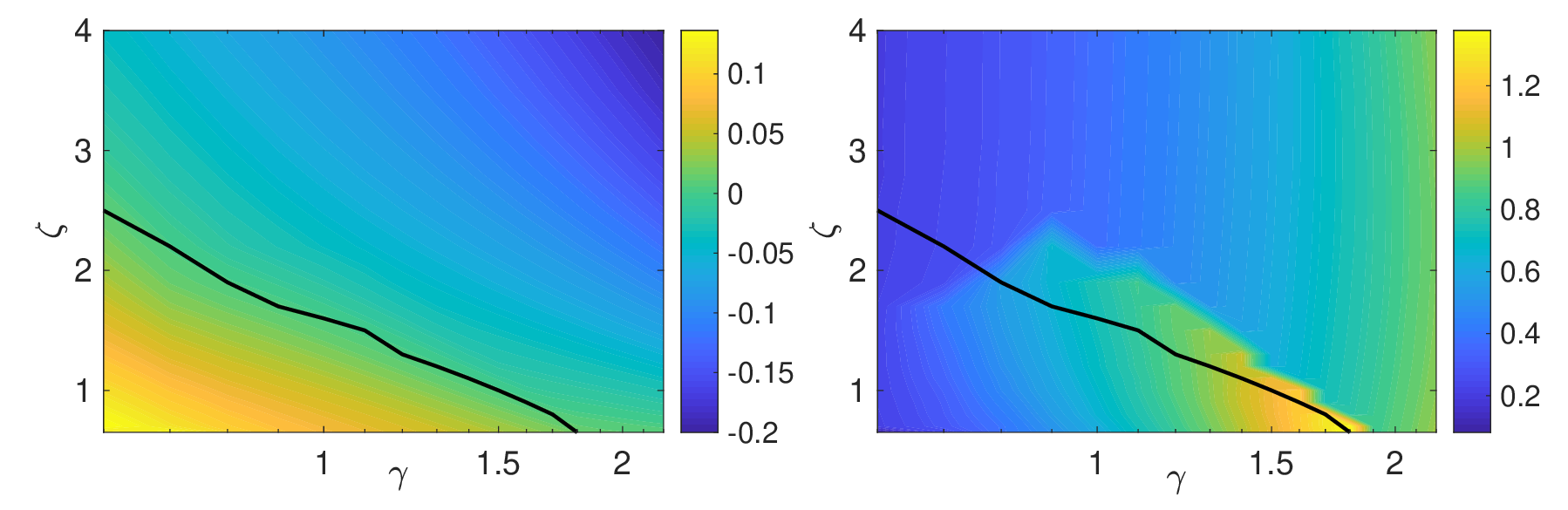}
    \caption{(Left) Linear growth rate($Re(\sigma)$) as a function of background shear and degree of confinement. (Right) Frequency($Im(\sigma)$) of the most unstable mode as a function of background shear and degree of confinement. The black curve is the neutral stability curve separating the stable region from the unstable region. }
    \label{fig:Growthrate}
\end{figure}

With that objective in mind, we first seek a 1D equilibrium solution by marching a 1D version of the reduced  equations Eqs.~\eqref{eqn:Phi_ROM}--\eqref{eqn:BC_ROM} and \eqref{eqn:adv_diff}~\cite{traverso_michelin_2022}. The existence of such a symmetric 1D solution  (Fig.~\ref{fig:1D_SS_ROM}) and its qualitative similarity with the 1D transient state of the full simulations (Fig.~\ref{fig:1Ddis2}) is a clear indication that the present reduced model is able to qualitatively capture strong wall polarisation and high particle density near the wall.   

In a second step, the full (2D) reduced equations are linearised around that 1D steady state:
\begin{equation}
\begin{split}
    &\Phi=\Phi_0(z) + \epsilon\Phi_1(y,z,t),\\
    &\bm{n}=\bm{n}_0(z) + \epsilon\bm{n}_1(y,z,t),\\
    &C=C_0(z) + \epsilon C_1(y,z,t),
\end{split}
\end{equation}
where  $0$ and $1$ subscripts refer to the steady 1D state and unsteady 2D perturbation, respectively. Linearising the governing equations at $O(\varepsilon)$, and assuming a normal mode decomposition in $(y,t)$ for the perturbation fields $\Phi_1,\bm{n}_1$  and $C_1$:
\begin{equation}\label{eqn:pert_var}
\Phi_1(y,z,t)=\tilde{\Phi}(z,k) e^{iky+\sigma t},\;\bm{n}(y,z,t)=\tilde{\bm{n}}(z,k) e^{iky+\sigma t}, \; C_1(y,z,t)=\tilde{C}(z,k) e^{iky+\sigma t}\cdot
\end{equation}
The linearised set of equations can be recast into an eigenvalue problem of the form 
\begin{equation}
\label{eqn:eigen}
    \mathcal{G}[\bm{x}_0]\cdot\tilde{\bm{x}}=\sigma\tilde{\bm{x}} 
\end{equation} 
where $\tilde{\bm{x}}$ is a column vector containing the perturbation amplitudes,  $(\tilde{\Phi},\tilde{n}_y,\tilde{n}_z,\tilde{C})$ 
with $\mathcal{G}[\bm{x}_0]$ a linear operator that depends on the 1D base state. The real and imaginary parts of the eigenvalue $\sigma$, namely $\mbox{Re}(\sigma)$ and $\mbox{Im}(\sigma)$ are respectively the growth rate and frequency of the perturbation.

This eigenvalue problem is discretized using a Gauss-Lobatto grid with $N+1$ points $(z^{(i)})_{1\leq i\leq N+1}$ across the channel width. The eigenvector $\tilde{\bm{x}}$ is now 
\begin{equation}
    \tilde{\bm{x}}=\big[\tilde{\Phi}(z^{(1)}),...,\tilde{\Phi}^{(N+1)},\tilde{n}_y^{(1)},...,\tilde{n}_y^{(N+1)},\tilde{n}_z^{(1)},...,\tilde{n}_z^{(N+1)},\tilde{C}^{(1)},...,\tilde{C}^{(N+1)}]
\end{equation} and the discretised linear operator $\mathcal{G}$ is now obtained from pseudo-spectral differential operators with modifications to include the boundary conditions corresponding to Eqs.\eqref{eqn:Phi_ROM}--\eqref{eqn:n_ROM}. Following Ref.~\cite{traverso_michelin_2022}, this eigenvalue problem, Eq.~\eqref{eqn:eigen} is solved numerically using MATLAB's algorithm based on the principle of minimized iterations~\cite{Arnoldi1951}. Eq.~\eqref{eqn:eigen} is solved for discrete values of $0\leq k\leq 2$  with a discrete step size of 0.01; the maximum value of $Re(\sigma)$ is reported in Fig.~\ref{fig:Growthrate}.

Figure~\ref{fig:Growthrate} (left) shows the variation of the growth rate of the least stable or most unstable mode (i.e. that with largest growth rate, $\mbox{Re}(\sigma)$), as a function of shear rate ($\gamma$) and confinement ($\zeta$).
The growth rate reduces with an increase of either background shear or degree of confinement demonstrating the stabilising effect of both mechanisms, already observed on the full simulation. 
Above the neutral curve, which corresponds to the parameters where the least stable mode is neutral ($\mbox{Re}(\sigma)=0$), the 1D fixed point is therefore stable (all eigenmodes have negative growth rate) with respect to 2D perturbations, an observation that is also consistent with the results of the full model that predict a $1D$ steady long-term dynamics for the larger values of $\gamma,\zeta$ (strong confinement or shear). 

For lower shear and/or confinement, there exists at least one unstable mode whose frequency $\mbox{Im}(\sigma)$ indicates the temporal nature of the dominant mode (oscillatory or monotoneous).
Fig.~\ref{fig:Growthrate} (right) shows that (i) $\mbox{Im}(\sigma)$ is non zero, so that the dominant mode is oscillatory in nature. Furthermore, $\mbox{Im}(\sigma)$ \change{is positive} and increases with shear, for the range of shear rate and confinement considered here. This observation is consistent with the results of the full simulations (Sec.~\ref{sec:Longterm_2D}) which noted the oscillatory effect introduced by an increasing shear, introduced by the periodic interaction of chemotactic aggregates located along each boundary as they are advected in opposite directions by the background shear. Indeed, the present minimalistic model includes such background advection ($\bm{u}\cdot \nabla_x \Phi$ in Eq.\ref{eqn:Phi_ROM}). 
Note however, that the full simulations predicted a \emph{steady} 2D regime at low shear rates, when the background shear is sufficiently weak for chemotaxis to be able to compensate the advection of opposite-wall aggregates. This discrepancy is somewhat not surprising, as it occurs in the low shear rate regime which we do not expect the present model to be able to reproduce/predict properly as one of the model's key assumption lies in its neglecting of all other contribution to the flow field than the background shear flow itself (see Appendix~\ref{sec:appendixA}).

\begin{figure}
    \includegraphics[width=\textwidth]{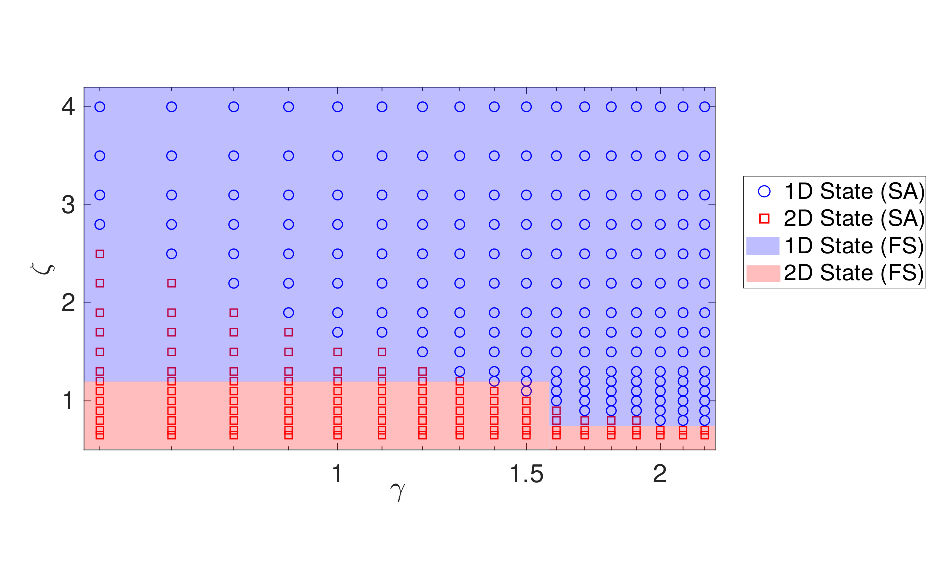}
    \caption{Comparison of the phase plot obtained via the linear stability analysis (\change{referred to as SA in the legend}) with the phase plot obtained using the full simulation (\change{referred to as FS in the legend})
    Red squares represent unstable 1D state while the blue circles represent the stable 1D state. 
    }
    \label{fig:Phase_ROM}
\end{figure}

\begin{figure}
    \centering
    \includegraphics[width=\textwidth]{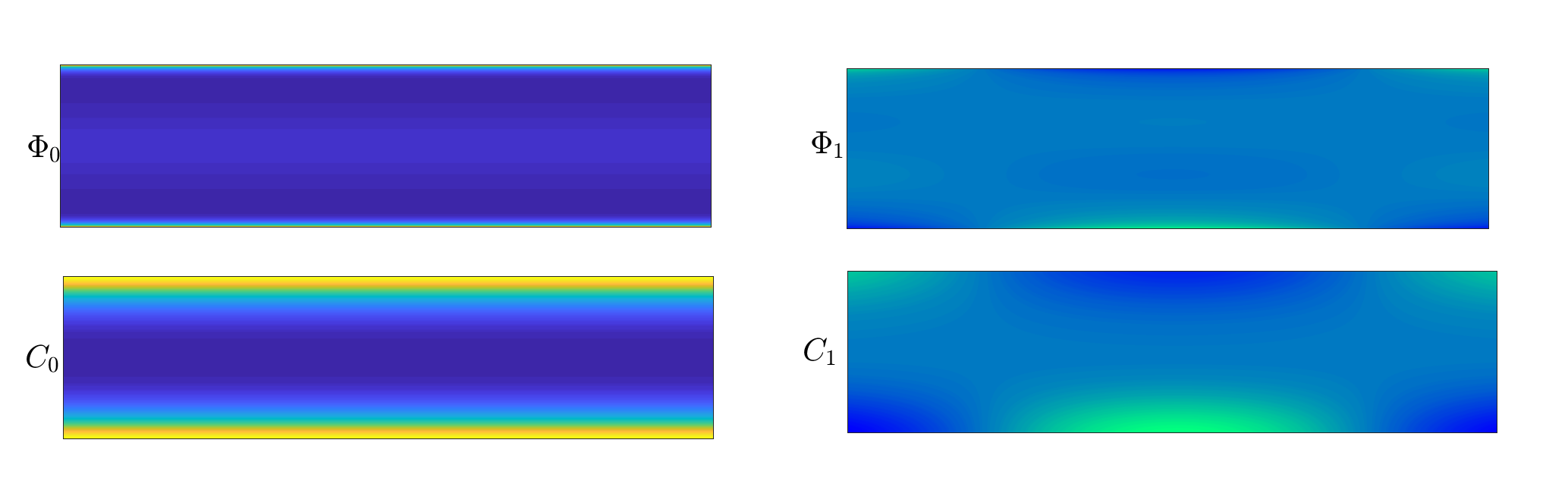}
    \caption{(Left) The steady state particle distribution and (right) real part of the eigenmodes
    of the particle density and solute distribution corresponding to the most unstable mode. }
    \label{fig:ROM_dist}
\end{figure}
Finally, based on $\mbox{Re}(\sigma)$ and $\mbox{Im}(\sigma)$ we plot a phase diagram similar to Fig.~\ref{fig:phase} as shown in Fig.~\ref{fig:Phase_ROM}.
In Fig.~\ref{fig:Phase_ROM} data points shown by red squares have a positive growth rate for the most unstable eigen mode with non-zero frequency. 
The most unstable eigen mode corresponds to asymmetric wall aggregates as shown in Fig.~\ref{fig:ROM_dist}, and therefore this region corresponds to 2D ($\mbox{Re}(\sigma)>0$) and unsteady($\mbox{Im}(\sigma)>0$) long term dynamics.     
Whereas, for data points marked with blue circles, the 1D fixed point is stable and the long-term solution is 1D steady state (despite $\mbox{Im}(\sigma)>0$ as the 1D fixed point is stable).  
This phase diagram qualitatively resembles with the phase diagram corresponding to the full simulations (Fig.~\ref{fig:phase}).
As expected, the match is particularly good in the strong shear region($\gamma \sim 2$) where the reduced order model correctly predicts the transition from the 2D unsteady state and 1D steady state at $\zeta\sim 0.8$.
 
To summarise, the reduced order equations show existence of a $y$-independent equillibrium state with strong particle and solute concentration at the walls. 
Streamwise perturbation of this 1D fixed point shows that asymmetric eigenmodes with aggregate formation on either wall exist for weak confinement.  
The imaginary part(representing the oscillation frequency) corresponding to the unstable eigen mode is non zero leading to 2D unsteady dynamics in the long term below the neutral stability curve(Fig.\ref{fig:Growthrate}). 
The linear perturbation analysis reveals that the 1D fixed point is stable for strong bakground shear and strong confinement resulting in a 1D steady state solution in the long term in agreement with full simulation results. 
Thus, this simple model is able to qualitatively capture complex self-organisation dynamics with sufficient accuracy.   

\section{Rheology of Active suspension}
\label{sec:sus_rheo}
\change{In this section, we analyze the rheological behaviour of the phoretic suspension as a result of the previously-discussed self-organisation. It is now well established that microswimmer suspensions, through the microscopic mechanical forcing they exert on their surroundings, can profoundly modify the macroscopic behaviour of the fluid and in particular its rheology~\cite{mussler2013,liu2019,Lopez2015}. We begin the analysis by defining an 'effective' viscosity of the suspension, based on the tangential stress exerted by the fluid and particles on the plate. The particles indeed modify the velocity field from a pure Couette flow and we thus analyze the flow patterns induced by the particles for the different states discussed in Sec.~\ref{sec:Longterm}. The flow organization results from multiple tightly linked factors, and in order to gain a better physical insight, a simplified model retaining the dominant phenomena is discussed in detail in Sec.~\ref{sec:RM_induced_flow}. Finally, the temporal variation of the effective viscosity for different states and the effect and shear and strength of confinement on effective viscosity is presented.} 


 \change{} 

\subsection{Defining an effective viscosity}
\label{sec:eff_visc}


\change{Viscosity is classically introduced as the ratio of the local stress and  strain rate in Newtonian fluids, however when the fluid or suspension shows a non-Newtonian behaviour, as for active suspensions, such an approach can become more difficult as the relative magnitude of stress and strain rate is expected to strongly depend upon the location considered.}

Alternatively, the viscosity can also be  defined based on classical \emph{global} results on parallel flows. This approach is commonly employed in effective viscosity measurements in Taylor-Couette devices for passive and active suspensions alike~\cite{Guazzelli2011,Rafai2010,Lopez2015}. This is also the point of view adopted for dilute phoretic suspensions in \change{pressure-driven} flows in Ref.~\cite{traverso_michelin_2022}, where the definition of an effective viscosity is based on the classical Poiseuille law relating the imposed pressure drop and flow rate within the channel. The advantage of such an approach is the particular relevance of pressure-driven pipe flows for different industrial, microfluidic or biomedical applications, but it overlooks the intrinsic non-uniformity of the imposed shear rate in such parabolic flow configuration. 

A similar approach is followed here on the simpler Couette-like flow configuration, where an anti-symmetric translation of the top and bottom boundaries results in a uniform shear stress distribution for Newtonian fluids. The effective viscosity can then be defined as the ratio of the force per unit area required to maintain the imposed translation of the boundaries, and compared to the Newtonian viscosity of the solvent obtained in the absence of the particles. More precisely, in the absence of particles, the force to apply on the plate is measured as $\hat{F}=\hat{\eta}\hat{\gamma} \hat{A}_s$, where $\hat{\eta} $ is the viscosity of the Newtonian fluid, $\hat{\gamma}=\hat{u}_w/\hat{H}$ is the imposed shear rate and $\hat{A}_s$ is the surface area considered. For a microswimmer suspension, we extend this definition by defining the suspension's effective viscosity $\hat{\eta}_e$ as $\hat{\eta}_{e}=\hat{F}/\hat{\gamma}\hat{A_s}$, with $\hat{F}$ now computed in the presence of the phoretic particles in terms of the total fluid stress tensor $\bm{\hat\Sigma}$ as 
\begin{equation}
\hat{F}= \int_{\hat{A}_s} \bm{\hat{n}_n} \cdot \bm{\hat{\Sigma}} \cdot \bm{\hat{t}_n} d\hat{A_s}\qquad \textrm{with   }\bm{\hat{\Sigma}}=-\hat{q}\bm{I}+\hat{\eta} \Big[\hat{\nabla}_x \bm{\hat{u}}+(\hat{\nabla}_x \bm{\hat{u}})^T \Big]+\bm{\hat{S}}
\label{eqn:force}
\end{equation}
where $\bm{\hat{n}_n}=-\bm{e}_z$ and $\bm{\hat{t}_n}=\bm{e}_y$ are the normal and tangential vectors to the upper plate. It should be noted here that the phoretic particles modify $\hat{F}$ (and $\hat{\eta}_e$) both directly via their active stresses, and indirectly via the viscous stresses exerted by the modified flow fields resulting from their self-organisation and cumulated forcing. The relative viscosity can then be defined as $\eta_r=\hat{\eta}_e/\hat{\eta}$.

Substituting the total stress field in Eq.~\eqref{eqn:force} and using the no penetration boundary condition at the plate results in 
\begin{equation}
\hat{F}=\int_{A_s} \bm{\hat{n}_n} \cdot \bm{\hat{\Sigma}} \cdot \bm{\hat{t}_n}d\hat{A}_s=\int_{A_s} \bigg( \hat{\eta} \frac{\partial \hat{u}_y}{\partial \hat{z}} +\bm{\hat{n}_n} \cdot \bm{\hat{S}} \cdot \bm{\hat{t}_n}\bigg)d\hat{A}_s.
\end{equation}
\change{As a result, the relative effective viscosity is given by,
\begin{equation}
\label{eqn:eff_visc}
  \eta_r=1+\frac{1}{A_s}\int_{A_s} \bigg( \frac{1}{\gamma }\frac{\partial u_{d,y}}{\partial y}+\frac{\bm{n_n}\cdot \bm{S}\cdot \bm{t_n}}{ \gamma }\bigg) dA_s.
\end{equation}
where $u_d$ and $\gamma$ are the disturbance velocity field and the shear rate respectively.}
It is important to note that in this formulation, the effect of the finite size of the particles and the influence of the resulting non-deformation stress are neglected. As a result, the effective viscosity of a passive suspension is equal to that of the pure solvent, i.e. thereby neglecting the Einstein's viscosity contribution to the suspension stress in this dilute limit~\cite{thomas1965}.

It is also noteworthy yet expected that the modification of the relative viscosity tends to zero in the limit where the imposed shear rate is large (in comparison with the diffusion of solute): as the externally imposed shear rate increases, the relative influence of the particles' active stress becomes negligible, and the particles behave similarly to passive particles as confirmed experimentally \cite{Rafai2010,Gachelin2013,Lopez2015}.

Its definition in Eq.~\eqref{eqn:eff_visc} identifies clearly two contributions to the effective viscosity, namely the Newtonian solvent stress resulting from the flow induced by the particles and the active stress exerted  by the particles directly on the wall. In an effort to elucidate more precisely the effect of the former, we first discuss the induced flow field generated for different shear forcing and confinement in Sec.~\ref{sec:Longterm} before considering to the global evolution of the effective viscosity in Sec.~\ref{sec:eta_t}.  

\subsection{Induced Flow}
\label{sec:induced}
\begin{figure}
    \includegraphics[width=0.9\textwidth]{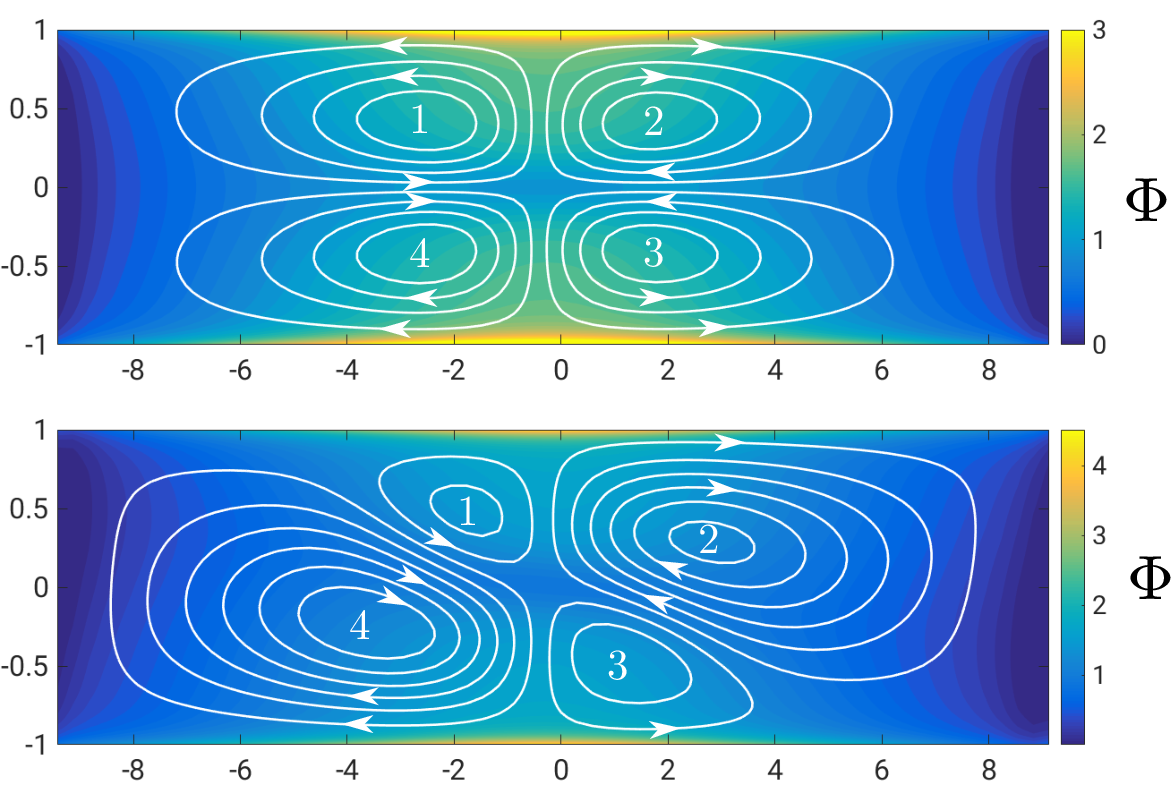}
    \caption{Streamlines of the induced flow for 2D steady state for $\gamma=0$(top) and $\gamma =0.0167$(bottom).}
    \label{fig:2DSS_str}
\end{figure}

\begin{figure}
  \includegraphics[width=0.8\textwidth]{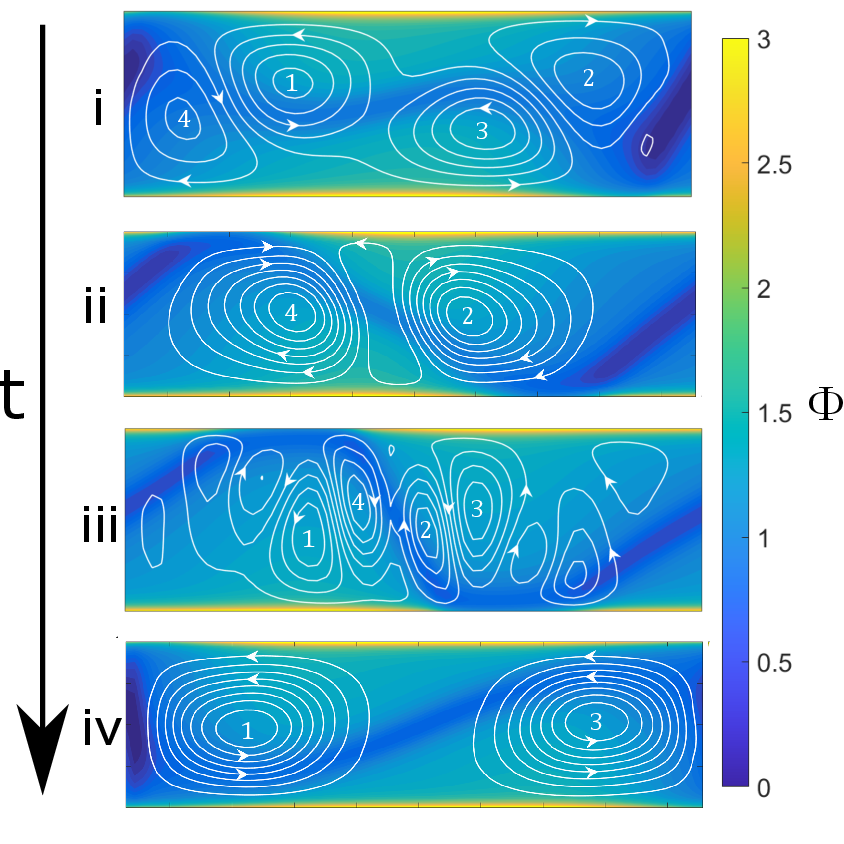}
  \caption{Evolution in time of particle density distribution and induced flow streamlines over a period of the unsteady regime, (i) $t=0$, (ii) $t\sim T/4$, (iii) $t\sim T/2$ and (iv) $t\sim 3T/4$ with 
  $\gamma=0.125,\zeta=1$. 
  }
  \label{fig:2Dstr}
\end{figure}
\begin{figure}
  \includegraphics[width=0.45\textwidth]{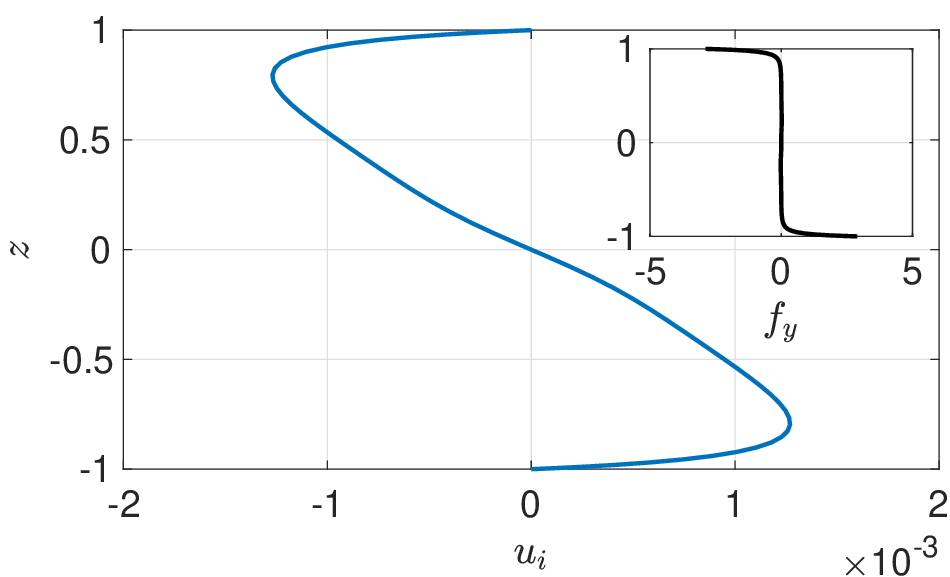}
\caption{(Left) Induced Flow field for the 1D symmetric flow for 1D steady state regime with $\gamma=0.125,\zeta=1.33$, inset shows the driving force due to active stress exerted by the particles. 
}
\label{fig:1Ddriv}
\end{figure}

The particle-driven disturbance flow is significantly influenced by the particles' self-organisation and we discuss here the characteristics of these induced flows for the different types of suspension dynamics observed at long times when varying confining and forcing conditions, as discussed in Sec.~\ref{sec:Longterm}. Stokes' equations are instantaneous; furthermore, the solute's relaxation is much faster than that of the particles: as a result, the induced flow at a given time essentially depends on the particles' distribution at that specific time only. In the following, we therefore discuss the induced flow field instantaneously, i.e. without considering the steady/unsteady nature of the self-organisation dynamics: in an unsteady regime, one expects to observe successively the different induced flows generated by the successive particles' organisation in time.

For weak confinement, and depending on the flow forcing, the active suspension self-organizes into regularly distributed wall aggregates along each wall either moving or stationary as discussed in Sec.~\ref{sec:Longterm_2D}. 
In the absence of a background flow, the aggregates are  placed symmetrically with particles oriented mainly toward the closest wall in response to the confinement-induced solute gradient (Fig.~\ref{fig:2DSS}). As a result the induced flow is also top-down and left-right symmetric   
and is characterized by two pairs of counter-rotating vortices (Fig.~\ref{fig:2DSS_str}), driven by the particles' aggregates with a dominant stagnation point flow toward  the wall driven by the aggregates. 

The introduction of a background shear  breaks the  symmetry in the particle organisation and transport  (see Sec.~\ref{sec:Longterm_2D}). 
Consequently, the induced flow also loses such top-down and left-right symmetries: for weak shear,  the induced flow still consists of four  vortices with one of the two pairs of co-rotating vortices becoming dominant over the other one which gradually disappears as shear is further increased Fig.~\ref{fig:2DSS_str}.
The rotation direction of the dominant and surviving vortex pair  strongly depends on the relative arrangement of the staggered vortices (Fig.~\ref{fig:2Dstr}).
For instance, if the top aggregate is displaced towards the right of the bottom aggregate, the counter-clockwise vortices in the first and third quadrant (1,3) are brought closer to and counteract each other thus forming a weaker pair, leaving the clockwise rotating vortices of the second and fourth quadrant (2,4) dominant. The reverse configuration is observed when the top aggregate is located slightly to the left of the bottom one (Fig.~\ref{fig:2Dstr}). As shear is further increased and the suspension's organisation becomes unsteady and periodic, the vortex arrangement can be understood similarly in terms of the successive aggregates' relative positions during a period, by exploiting the instantaneous nature of the Stokes problem.  
Such vortex flows are reminiscent of flows observed for bacterial suspensions \cite{wioland2016,neef2014vort,henshaw2023dyn} and in other active systems \cite{shendruk2017dancing}.
A more detailed explanation of the induced flow is discussed in the next  subsection, in particular the link between local polarization, the direction of the concentration gradient and the induced flow.

Strong confinement stabilizes the chemotactic instability resulting in 1D particle distribution with high particle density near the channel walls as discussed in Sec.~\ref{sec:Longterm1D}. The particles are strongly polarised toward the wall but can be slightly tilted by the imposed shear, resulting in a net horizontal fluid forcing near the wall regions. 
Such 1D induced flow closely resembles induced flow already reported in experimental and numerical studies on microswimmer suspensions as well \cite{theillard2017,wioland2016,Lushi2014}. 

\subsection{A simplified model for the induced flows}
\label{sec:RM_induced_flow}


In an effort to provide a more intuitive insight into the role of particle distribution, local polarisation and solute concentration gradient in the establishment of the induced flows, a qualitative form of the fluid forcing induced by the particles is presented in this section. To this end, qualitative observations are made on the numerical results for the parameter values and ranges considered here.

While the magnitude of the self-induced stresslet $\bm{S_s}$ is intrinsic and fixed for each particle, that of the stresslet $\bm{S_e}$ induced by external concentration gradients depends on the local concentration gradient,Eq.~\eqref{eqn:stress_ex}. Therefore depending on the local solute distribution arrangement, each particle will either behave as a net pusher or a net puller. 
The no-flux condition at the wall for the solute concentration together with solute diffusion ensures that $| \nabla_x C|\sim O(1)$ everywhere and at all times (see Fig.~\ref{fig:gradc} and appendix~\ref{sec:AppendixB}).
For the specific  values chosen here for $\xi_t$, $\xi_r$ and $u_0$, the strength of the pusher contribution is therefore almost twice that of the puller contribution, which is further confirmed by noting the similarity in flow patterns (see Fig.~\ref{fig:complete_and_pusher} and appendix~\ref{sec:AppendixB})) obtained for the full forcing or using solely the pusher contribution (i.e. ignoring the externally-induced stresslet). Note that such similarity is also observed throughout the simulation.
This suggests that the phoretic particles considered here behave as net pushers with modified stress intensities $\sigma_m\sim \alpha_s+\alpha_e$ ($\sigma_m<0$).    

Furthermore, as a result of the relatively strong concentration gradients and chemotactic behaviour of the particles, the suspension is strongly polarised (i.e. $|\bm{n}|$ roughly close to 1, see Fig.~\ref{fig:gradc}) in particular close to the wall where solute and particles accumulate: locally, most particles share the same orientation which is close to that of the local solute concentration gradient. As a first approximation, it is therefore possible to consider that ${\Psi(\bm{x},\mathbf{p},t)\sim\Phi(\bm{x},t)\delta(\bm{p}-\bm{n}(\bm{x},t))}$ with $\bm{n}\approx\nabla_x C/|\nabla_x C|$, which significantly simplifies the description of the suspension. In particular, the average active stress field is now simply given by
\begin{equation}
\label{eqn:approx_stress}
 \langle \bm{S} \rangle = \int_\Omega \Psi(\bm{x},\bm{p})\bm{S(\bm{p})}d\bm{p} \approx \bm{S(\bm{n})} \int_\Omega \Psi(\bm{x,p}) d\bm{p}=\sigma_m\Phi(\bm{x})\left(\bm{nn}-\frac{\bm{I}}{2}\right).
\end{equation}
The fluid forcing is then 
\begin{equation}
\bm{f}=\nabla_x\cdot\langle\bm{S}\rangle=\bm{S(n)}\cdot \nabla_x \Phi+\Phi(\nabla_x\cdot \bm{S(n)}).
\label{eqn:forcing}
\end{equation} 
Both terms of Eq.~\eqref{eqn:forcing} in fact roughly provide similar forcing throughout the domain (with different magnitudes) as seen in Fig.~\ref{fig:stress_div}. As a result, retaining one of the two terms of Eq.~\eqref{eqn:forcing} with a corrected amplitude ($\nu>0$ in this case based on Fig.~\ref{fig:stress_div} in Appendix.~\ref{sec:AppendixB}). Thus,

\begin{equation}
\bm{f} = \nu \bm{S}\cdot \nabla_x\Phi
\end{equation}
 further  simplifies the problem's description and treatment without qualitatively changing the effect of active forcing. The exact value of $\nu$ can be determined by taking the ratio of the two components; we however note that the precise chosen value does not modify the conclusion qualitatively.
 
We now employ the relation to quantitatively understand the induced flow for 1D and 2D regimes.

\subsubsection{1D regime}
\label{sec:induced_flow_1D}

As the particle density varies only in the vertical direction, the fluid forcing simplifies to 
\begin{equation}
\bm{f}=\nu\sigma_m \bigg(\bm{nn}-\frac{\bm{I}}{2}\bigg)\cdot \frac{d\Phi}{dz} \bm{e_z}
\end{equation}

We are particularly interested in the horizontal component of this forcing, namely $f_y=\nu\sigma_mn_zn_y\frac{\mathrm{d}\Phi}{\mathrm{d}z}$, as it is responsible for the emergence of the induced flow observed in Fig.~\ref{fig:1Ddriv} -- the $y$-independent vertical forcing simply modifies the pressure distribution across the channel. Here $\nu>0$ (see Fig.~\ref{fig:stress_div}) and $\sigma_m<0$ 
and emergence of horizontal forcing is therefore tightly linked to horizontal polarization, away from the local (mostly vertical) solute gradient. Such a  tilt of the particles is caused by the background shear flow which  rotates the spherical particles in the clockwise direction throughout the channel. In a steady regime, this hydrodynamic torque is balanced by the chemotactic one that tends to bring the particles back to a vertical orientation; the particles thus maintain a slight clockwise tilt, i.e. $n_y>0$  \change{($n_y<0$)} in the upper \change{(resp. lower)} half  of the channel. In that region, the concentration gradient and particles' vertical polarisation are directed toward the upper \change{(resp. lower)} wall, $\mathrm{d}\Phi/\mathrm{d}z,n_z>0$ \change{(resp. $\mathrm{d}\Phi/\mathrm{d}z,n_z<0)$} and $\sigma_m<0$ so that $f_y<0$ \change{(resp. $ f_y>0$)} and the induced forcing acts against the background flow. 


The flow forcing by the particles is maximum at the walls where a no-slip condition is enforced, and thus results in a maximum magnitude of the induced flow field slightly away from the no-slip walls (Fig.\ref{fig:1Ddriv}).

\subsubsection{2D regimes}
\begin{figure}
\includegraphics[width=0.9\textwidth]{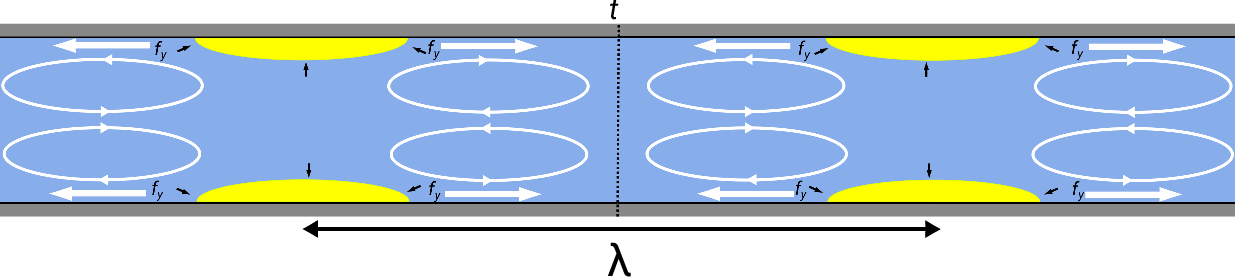}
\caption{Illustration of the fluid flow for 2D symmetric case. The white arrows show the direction of horizontal fluid forcing, and the black arrow shows particle polarization. The fluid forcing changes direction at a point t equidistant from the aggregates in the same wall due to a change in horizontal polarization. }
\label{fig:2D_vortices}
\end{figure}
\begin{figure}
\includegraphics[width=0.9\textwidth]{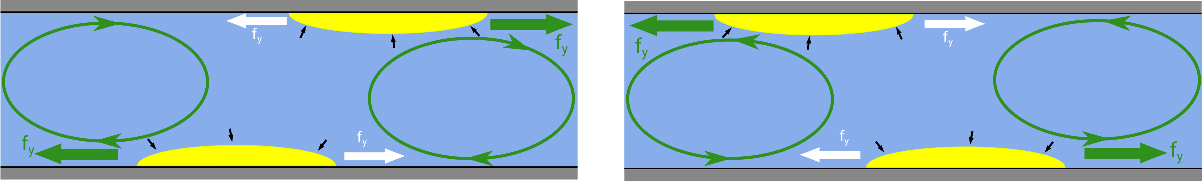}
\caption{Schematic showing asymmetrical horizontal forcing which leads to two vortices with the same direction of rotation. Green arrows show the relative dominant flow forcing resulting in the two vortex flows shown in green. The relative increase in the flow forcing is due to the orientation bias created to due to the presence of aggregate on the opposite wall.}
\label{fig:2D_2vortices}
\end{figure}

We now turn to the 2D suspension dynamics and first consider the symmetric and steady flow induced in the absence of any background flow as discussed in Sec.~\ref{sec:Longterm_2D}.
The particles are positively aligned along the solute gradient which roughly follows the gradient $\nabla_x\Phi$ in particle density. The particles are net pushers and exert an extensile forcing along their direction, i.e. toward and away from the aggregates. As the particle concentration increases toward the aggregates, the forcing by each particle toward the accumulation region is counteracted by a stronger forcing in the opposite direction by the particles located in front of it. As a result, the net flow forcing by the particles is oriented against their polarisation and away from the chemotactic aggregates. Consequently, particles present on the right (left) of aggregate induce a flow in positive (negative) $y$-direction, leading to a pair of counter-rotating vortices oriented as illustrated on Fig.~\ref{fig:2D_vortices}, and by mass conservation, a vertical flow pumping toward each aggregate and recirculation into the four-cell structure described in Sec.~\ref{sec:induced}. 

The same arguments remain applicable for non-symmetric chemotactic aggregates. The top-down symmetry is now broken resulting in an asymmetry of the horizontal forcing (Fig.~\ref{fig:2D_2vortices}). Considering for example the configuration where the top aggregate is positioned on the right of its bottom counterpart, particle horizontal polarisation around the top aggregates is now weaker on the left, where it is perturbed by the closer presence of the bottom aggregate, than on the right side, resulting in a stronger forcing by the latter that drives the  dominant clockwise vortex below. Similar arguments can be followed to rationalise the dominance of a pair of counter-clockwise vortices when the top aggregate is located on the left of the bottom one.

\subsection{Time evolution of the effective viscosity}
\label{sec:eta_t}

\begin{figure}
\includegraphics[width=\textwidth]{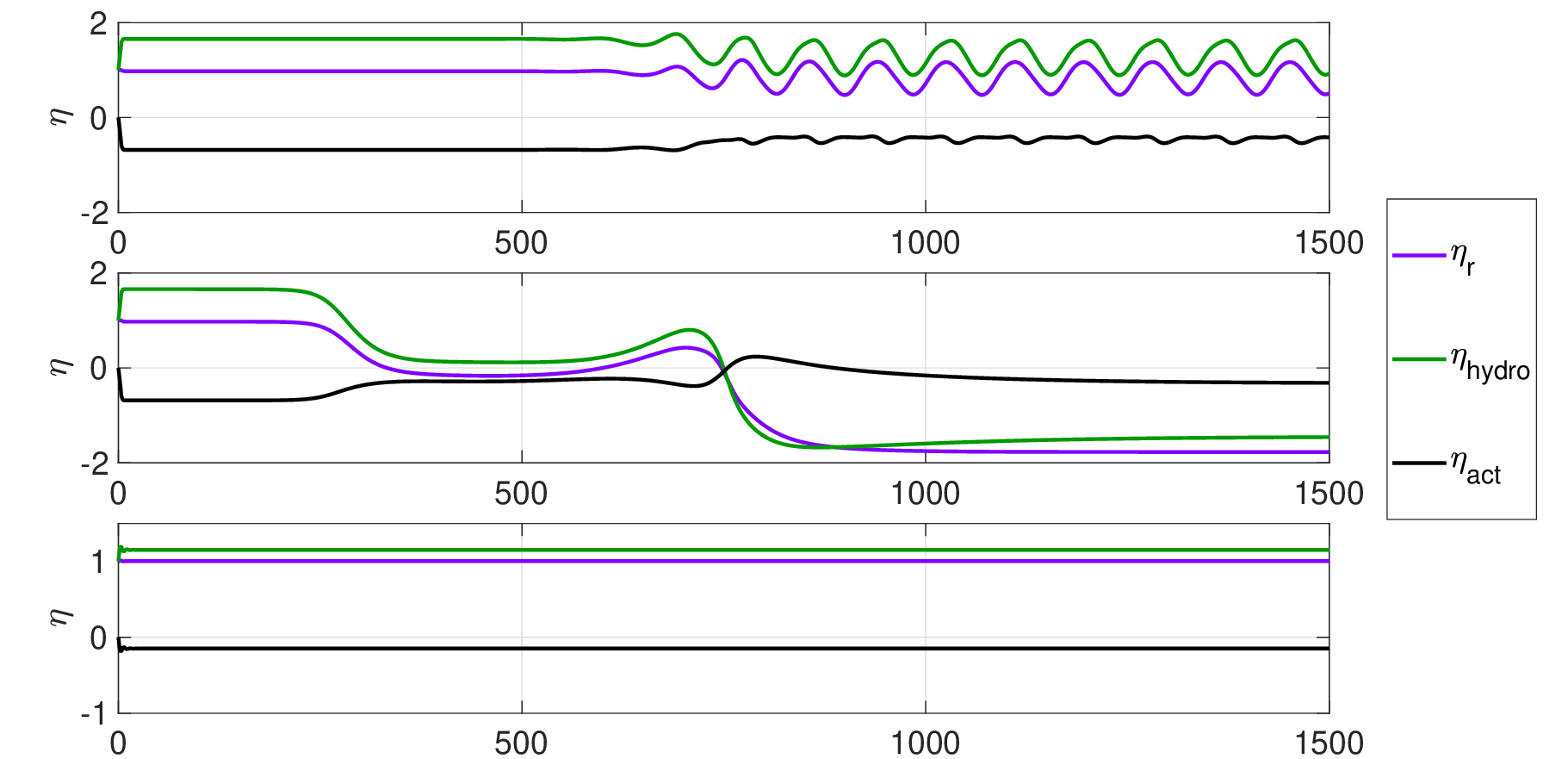}
\caption{Time evolution of effective viscosity($\eta_r=\eta_{hydro}+\eta_{act}$) for 2D unsteady (top) state and 2D steady state(centre) and 1D steady state (bottom) for parametric values $\gamma=0.125,\zeta=1$(top), $\gamma =0.025,\zeta=1$(centre) and $\gamma =2,\zeta=1$. }
\label{fig:visc}
\end{figure}

\begin{figure}[t]
 \includegraphics[width=0.45\textwidth]{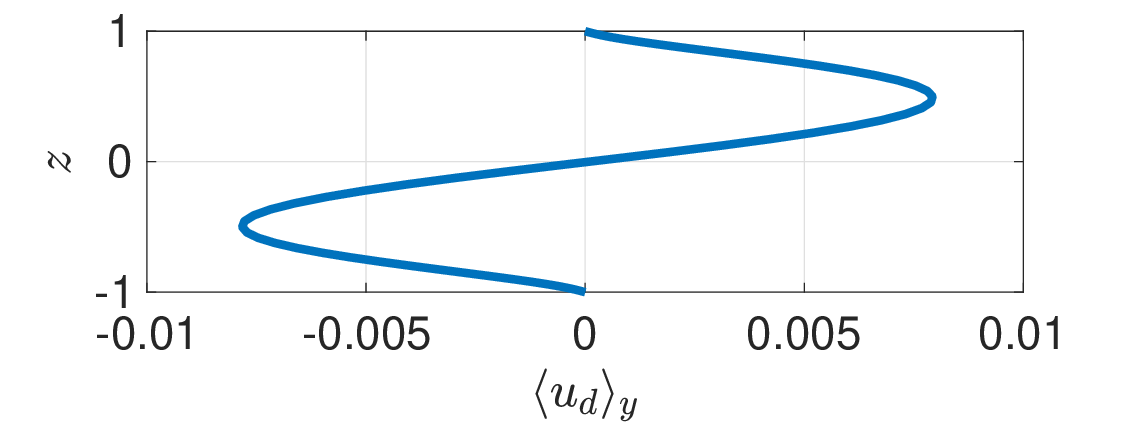}
  \includegraphics[width=0.45\textwidth]{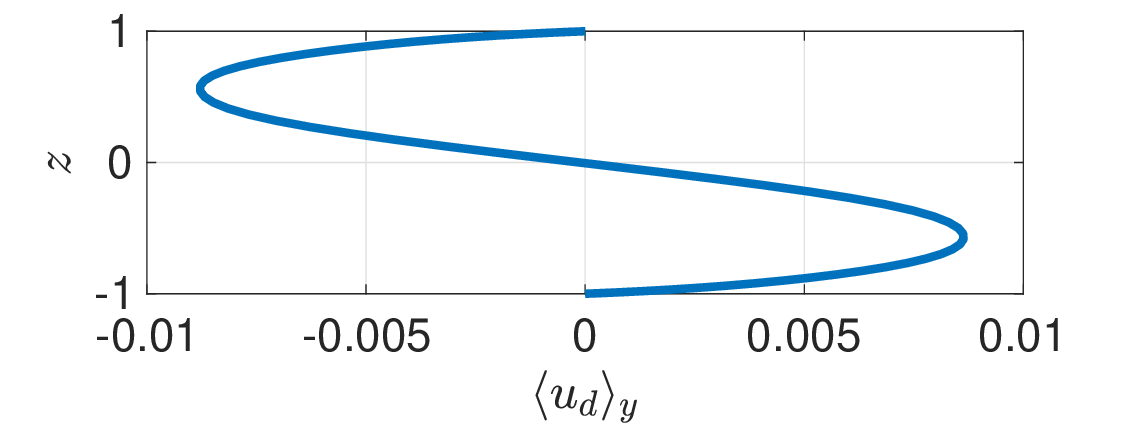}
  \caption{$\langle u_d\rangle_y$ (average disturbance velocity profile) for the 2D particle distribution with top aggregate displaced right (left) of bottom aggregate and (right) top aggregate displaced to the left of bottom aggregate for $\gamma=0.125,\zeta=1$ (Unsteady regime). }
  \label{fig:2Dflow2}
\end{figure}

\begin{figure}[t]
    \includegraphics[width=0.9\textwidth]{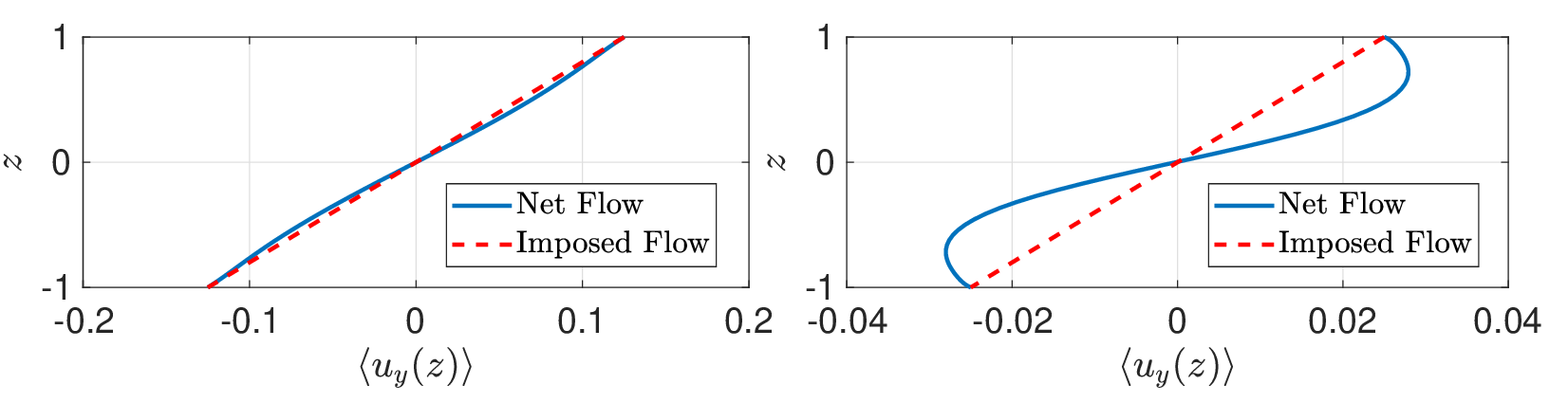}
    \caption{Net flow(averaged along the flow direction) compared to the imposed flow for shear rates 0.125 (unsteady regime) and 0.025 (steady regime) and $\zeta=1$ at viscosity minima for the 2D unsteady state.}
    \label{fig:flow_compare}
\end{figure}
Having understood the flow field forced by the particles in the different regimes, it is now possible to consider the  modified force exerted by the suspension on the moving plate and compute the effective viscosity of the suspension in this Couette geometry as defined in Sec.~\ref{sec:eff_visc}. The time evolution of viscosity is directly related to the suspension's self-organisation and its steady/unsteady evolution is a clear reflection of the steady/unsteady nature of the particle distribution. 

\begin{figure}
\includegraphics[width=0.9\textwidth]{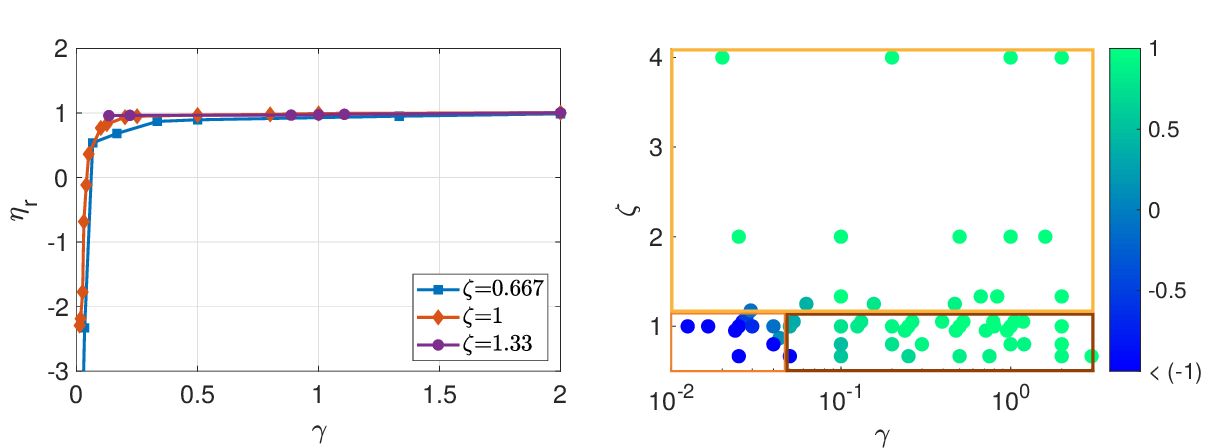}
\caption{(Left)Variation of effective viscosity with respect to background shear rate for different degrees of confinement($\zeta$). (Right) Long term effective viscosity$(\eta_r)$ on shear rate($\gamma$) - confinement space($\zeta$). The range of the colour axis is modified such that the effective viscosity below $-1$ is all coloured identically as the data points are highly skewed for weak shear rates and weak confinement. Rough boundaries are drawn for the different long term regimes for respective shear rate and strength of confinement. Golden boundary indicates long term 1D steady state, orange boundary indicate long term 2D steady state and brown boundary indicate long term 2D unsteady state.}
\label{fig:eta}
\end{figure}

In all simulations,  $\eta_r=1$ initially in all cases \change{(there is no net induced flow for an isotropic and uniform suspension)}. A weak reduction of the effective viscosity is observed during the transient 1D state, but the decomposition of the suspension forcing on the plate into a hydrodynamic part (the shear force resulting from both the imposed and induced flows) and an active part (stress exerted directly by the particles on the plate) shows that both effects are of appreciable amplitude and act  in opposite directions, with the hydrodynamic and active contributions respectively enhancing and reducing the total force to apply on the plate and the effective viscosity. The former is the result of the induced flow counteracting the background shear flow resulting from the rightward motion of the plate, thus enhancing the velocity gradient at the wall and resulting shear force. The direct active force exerted by the particles on the top wall can be written as ${f_w=\bm{n_n}\cdot \bm{S}\cdot \bm{t_n}\approx -\sigma_mn_yn_z\Phi}$, with $\bm{n_n}=-\bm{e}_z$ and $\bm{t_n}=\bm{e}_y$  the unit normal and tangent vectors at the wall. Here, $\sigma_m<0$ 
and $n_yn_z>0$ at both the walls as discussed previously, resulting in a net force pushing the plate in the flow direction, thus reducing the effective viscosity. 
\change{As the 1D transition state is stable for strong confinement, the viscosity remains at a constant value throughout the simulation (Fig.~\ref{fig:visc}). This corresponds to the region marked with yellow boundaries in Fig.~\ref{fig:eta} (right) and the plateau region in Fig.~\ref{fig:eta} (left) which correspond to high shear.  }

For weak confinement and strong shear forcing, the long-term suspension's response and effective viscosity are unsteady but periodic  (Fig.~\ref{fig:visc}). The oscillation of viscosity can be understood as the result of the changing directions of rotation of the vortex cells identified in the unsteady 2D regime (see Sec.~\ref{sec:Longterm_2D}).  
When the top aggregate is located to the right of the bottom one, a system of clockwise vortices is generated that tends to entrain the top plate in its direction of motion, thus reducing the velocity gradient and shear stress at the wall, or reducing the effective viscosity. Instead, the emergence of a counter-rotating vortex system entrains the plate in the opposite direction, thus enhancing the viscous shear stress and effective viscosity at the wall. 

\change{For weak shear rates ($\gamma\ll u_0$),  the suspension's dynamics is steady, with chemotactic aggregates on the top wall shifted to the right (see Sec.~\ref{sec:Longterm_2D}), resulting in 
a clockwise dominant vortex system is observed which reduces the shear gradient at the walls. Consequently, reduction, in fact reversal, of the hydrodynamic forcing on the plate, leads to a net negative viscosity (Fig.~\ref{fig:visc}). This regime corresponds to maximum viscosity reduction due to i) clockwise rotating vortices and ii) low shear which enhances the relative contribution of active stress. This regime is represented with orange boundary in Fig.~\ref{fig:eta}.}  



\section{Conclusions}
\label{sec:conc}
Based on a kinetic model, this work analysed numerically and theoretically the self-organisation dynamics of a two-dimensional dilute suspension of auto-phoretic particles under the dual forcing of confinement and of a background shear (Couette flow). \change{In comparison with earlier studies~\cite{traverso_michelin_2022}, this setup allows a more precise investigation of the relative and coupled effects of shear and confinement by releasing the correlation of strong shear and strong confinement present in pressure-driven flows. The results presented here further investigate the whole range of confinement intensities to bridge the gap between confinement-driven dynamics and the spontaneous bulk one.}  The dynamic response of the suspension provides some important qualitative and quantitative insights on the rheological behaviour of such chemotactic active suspensions.

Starting from a perturbed uniform and isotropic distribution of Janus phoretic particles within the channel, a rapid development of a one-dimensional (cross-channel) distribution is a common feature for the range of confinement and shear rate intensities considered in this work, and results from the swimming particles' accumulation in the immediate vicinity of the bounding walls. At longer times, its persistence depends on the competition of this effective wall attraction with chemotaxis. For sufficiently strong confinement, this  1D steady state remains stable to streamwise perturbations and thus observed at large times for small channel widths. However, when the bounding walls are too far apart, streamwise perturbations destabilize this 1D regime as a result of the chemotactic instability~\cite{Traverso2020}, which results in the formation of particle aggregates on the walls. These aggregates are transported by the background flow in opposite directions along each wall. When the shear rate is low enough, the chemotactic attraction of opposite-wall aggregates is sufficient to maintain a steady two-dimensional regime with offset positions of the particle clusters across the channel. Beyond a critical shear rate, chemotaxis attraction cannot compete with particle and solute transport by the flow resulting in a periodic  two-dimensional dynamics of the system, which is asymmetric in time as a result of the retarded chemotaxis response. A simple reduced model based on the particle density and polarization is proposed and shown to be sufficient to capture the flow forcing and the induced flow qualitatively.

In a second step, the hydrodynamic forcing exerted by the particles on the surrounding flow is computed to analyse the dynamical response and resistance exerted by the suspension on the moving walls, providing insight into the effective viscosity (i.e. force response to a given shear rate). The modification from the solvent viscosity is two-fold, resulting both from the active stresses exerted by the particles which modify the velocity gradients (and shear force) at the walls, and from the direct forcing exerted by the particles on the walls. 

In agreement with now-classical rheological behaviour of bacterial suspensions~\cite{Lopez2015,Rafai2010}, this work shows that the modification in effective viscosity is largest for weak background shear: active stresses are then relatively stronger. In contrast for large imposed shear rates, the background forcing dominates the flow dynamics and forces, and particle dynamics are essentially similar to that of passive colloids. Consequently, the suspension maintains a Newtonian behaviour at larger shear rates (Fig.~\ref{fig:eta}, left).

For low shear rates, the self-organisation of the confined suspension is directly responsible of the complex non-Newtonian behaviour of the suspension, and is characterised by significant reduction in the   effective viscosity as a result of the active forcing of the particle. This forcing results from the competiting surface-driven flows generated by the particles in response to their chemical activity and the phoretic forcing of the suspension's solute distribution.

The sensitivity of the particle distribution to the relative effects of convection and background shear, and the dual response of the particles to hydrodynamic and chemical forcing opens up the possibility \change{ to influence  the suspension's rheological properties indirectly}. The self-organisation of a dual-response suspension has indeed already been reported for other systems using  magnetic~\cite{snezhko2011}, electric~\cite{Park2011}, chemical~\cite{hong2007}, or optical\cite{palacci2013,garcia2013} forcing. A similar control of the suspension would open up some particularly interesting routes for application and should be investigated in future studies.
\section*{Acknowledgements}
 P.V. is grateful for the fruitful discussions with Dr T. Traverso and Dr N. Desai. This work was supported by the European Research Council (ERC) under the European Union's
Horizon 2020 research and innovation program (Grant Agreement No. 714027 to S.M.).
\appendix
\section{Reduced order equations}
\label{sec:appendixA}
We outline here the derivation of the reduced order equations, which closely follows that in Ref.~\cite{traverso_michelin_2022} to which the reader is referred to for more details. 

The $\bm{p}$-dependance of the probability density function $\Psi$ can be decomposed onto spherical harmonics of successive orders, thereby decomposing $\Psi$ as an infinite sum of orientation moments~\cite{Saintillan2013}. Each moment corresponds to a physical quantity, and contributes to the characterisation of the variability in particle orientation. For instance, the zeroth order moment corresponds to the local particle density $\Phi$, the first order moment  to the local average orientation or polarisation of particles $\bm{n}$, the second moment correspond to the nematic order, and so on. This expansion is truncated here after the first two moments, resulting in
\begin{equation}
\Psi(\bm{x,p},t)=\frac{1}{2\pi}\Phi(\bm{x},t)+\frac{1}{\pi}\bm{p}\cdot\bm{n}(\bm{x},t).
\end{equation}  
Taking successive moments of the Smoluchowski equation, Eq.~\ref{eqn:psi}, with respect to $\bm{p}$ provides the equations of evolution for the particle concentration and polarisation. Note that classically, a closure relationship is needed as directional self-propulsion introduces a forcing of each moment by higher order ones; following, Ref.~\cite{traverso_michelin_2022} the nematic ordering is thus represented as  
\begin{equation}
\bm{Q}(\bm{x},t)=\langle \bm{pp}-\frac{\bm{I}}{2}\rangle\approx \frac{\Phi\bm{I}}{2}.
\end{equation} 
As the dynamics of the suspension can be qualitatively understood without including the effect of the induced flows on the particles' transport, we further disregard such contributions so that the flow field used in the evaluation of the particles' transport is simply the background shear flow. This essentially decouples the Stokes equations from the particle distribution dynamics and results in the following evolution equations for $\Phi$ and $\bm{n}$:  
    \begin{equation}
    \label{eqn:Phi_ROM}
   \frac{\partial \Phi}{\partial t}+\bm{u}\cdot \nabla_x \Phi=  -u_0\nabla_x \cdot \bm{n} -\frac{\xi_t}{\zeta}\big[\nabla_x C \cdot \nabla_x \Phi +\Phi \nabla^2_xC\big]+d_x\nabla^2_x\Phi 
\end{equation}
\begin{equation}
\label{eqn:n_ROM}
\begin{split}
   \frac{\partial \bm{n}}{\partial t}+\bm{u}\cdot \nabla_x \bm{n}= & -\frac{u_0}{2}\nabla_x \Phi -\frac{\xi_t}{\zeta}\big[\nabla_x C \cdot \big(\nabla_x \bm{n}\big)^T+\bm{n}\nabla^2_xC\big]+\frac{\xi_r \Phi\nabla_x C}{2\rho\zeta} +
    d_x\nabla_x\cdot\big(\nabla_x \bm{n} \big)^T \\& -d_p \bm{n}+\frac{\gamma}{2} \bm{n} \cdot\big(e_z e_y-e_ye_z\big) 
\end{split}
\end{equation}
In Equation~\eqref{eqn:Phi_ROM}, the successive terms on the right hand side correspond respectively to self-propulsion, phoretic drift and translational diffusion of the particles, respectively, while in Eq.~\eqref{eqn:n_ROM}, the successive forcing terms can be identified as self-propulsion, phoretic drift, chemotaxis, translational and rotational diffusions, and reorientation by the background vorticity.

The boundary conditions are  evaluated similarly from Eq.~\eqref{eqn:no_flux_psi} as
\begin{equation}
\label{eqn:BC_ROM}
u_0n_z=d_x\frac{\partial \Phi}{\partial z}, \hspace{0.05\textwidth} \frac{\partial n_y}{\partial z}=0, \hspace{0.05\textwidth} u_0\Phi=d_x\frac{\partial n_z}{\partial z}\qquad  \textrm{at}\quad z=\pm1.
\end{equation} 
The solute concentration evolution equation and corresponding boundary conditions remain unchanged, Eq.~\eqref{eqn:adv_diff}. 

\section{Simplification of the fluid forcing}
\label{sec:AppendixB}
In this appendix, we revisit and justify the assumptions made in Sec.~\ref{sec:RM_induced_flow} to simplify the flow forcing.

\begin{figure}
  \includegraphics[width=0.7\textwidth]{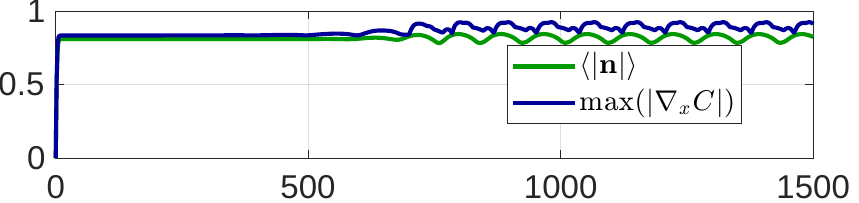}
  \caption{Time evolution of spatial mean polarization magnitude and  maximum $|\nabla_x C|$ for weak confinement case. The plateau region corresponds to the 1D transient state and the long-term periodic behaviour corresponds to the long-term unsteady 2D state. The plots are for $\gamma=0.125$, $\zeta=1$. 
  }
  \label{fig:gradc}
  \end{figure}
  

\begin{figure}
  \includegraphics[width=0.9\textwidth]{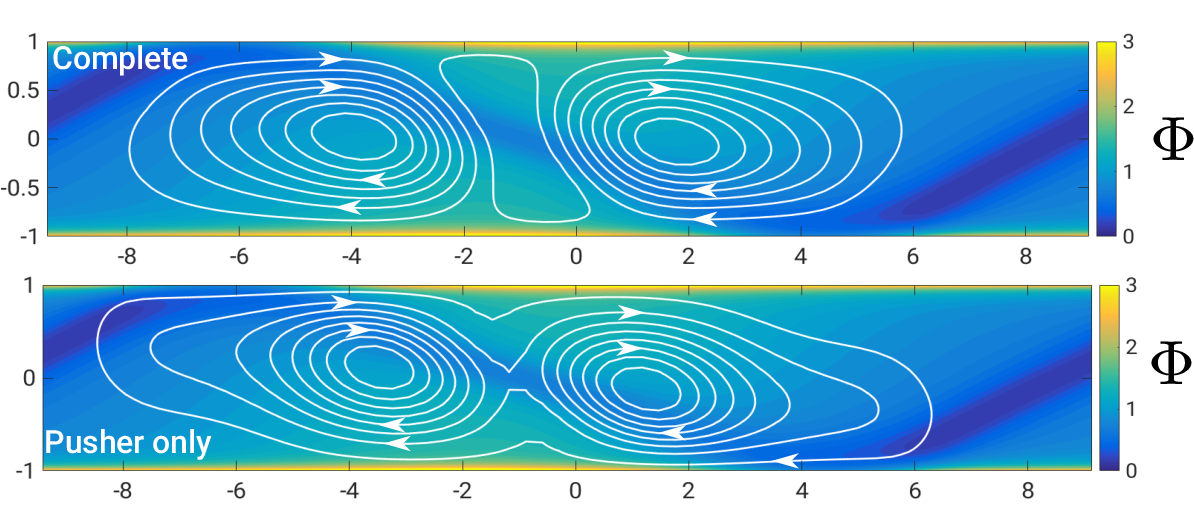}
  \caption{Comparison between disturbance velocity field due (top)total active stress and (bottom)only the pusher signature of active stress for $\gamma=0.125,\zeta=1,t=1325$. The colour bar represents the particle density in the domain. }
  \label{fig:complete_and_pusher}
  \end{figure}
  
First, we approximate that the particles in the system are locally aligned completely. This assumption is supported by the observation that the spatial mean in particle polarization ($\langle \bm{|n|}\rangle$) is close to 1, as depicted in Fig.~\ref{fig:gradc}. This suggests that the pusher ($\mathbf{S_s}$) and puller contribution ($\mathbf{S_e}$) directly compete, resulting in the particle behaving as a net pusher/puller. We approximate the net behaviour of the particles as pushers based on the maximum strength of the concentration gradient, which remains of $O(1)$ (as shown in Fig.~\ref{fig:gradc}.), suggesting that the pusher contribution is the dominant contribution. As a result, the fluid forcing can be approximated as a product of local particle density and the pusher contribution $\mathbf{S_s}$ in Eq.~\eqref{eqn:approx_stress}.


The approximation of pusher behaviour for each particle is further validated by observing the close similarity in the induced flows by including both the contributions (top) and only the pusher contribution (bottom) in Fig.~\ref{fig:complete_and_pusher}. The background contour plot in Fig.~\ref{fig:complete_and_pusher} shows the domain's particle density distribution ($\Phi$). 
  
  \begin{figure}
  \includegraphics[width=0.75\textwidth]{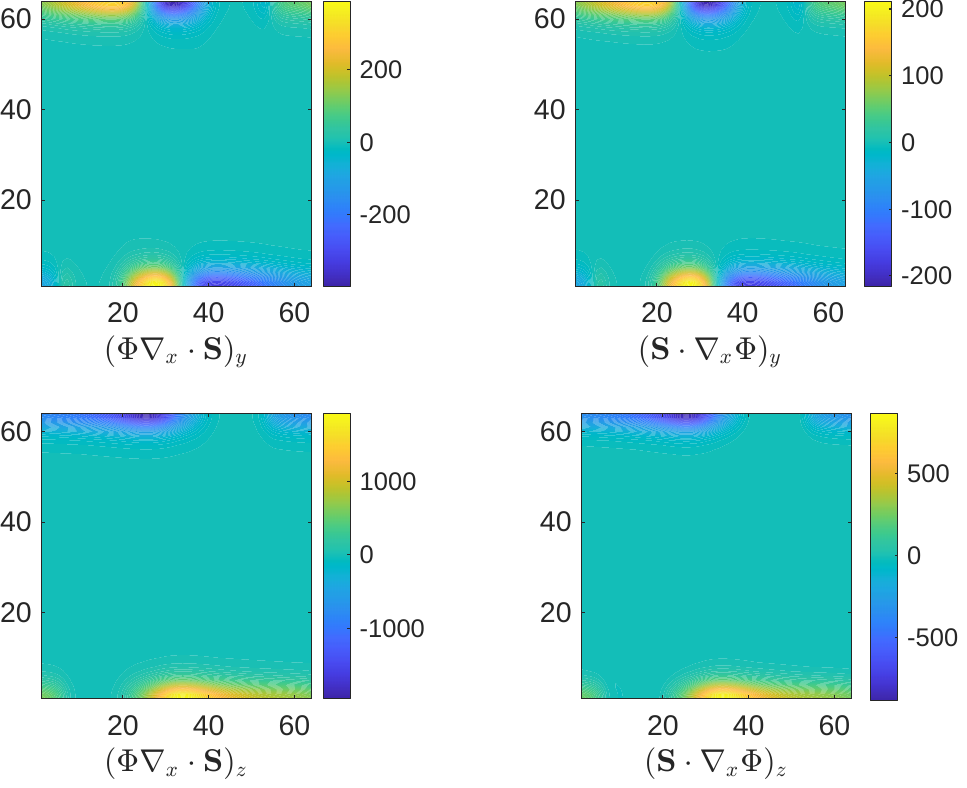}
  \caption{Horizontal and vertical fluid forcing for the two terms of Eq.~\ref{eqn:forcing} for $\gamma=0.125,\zeta=1,t=1500$. The forcing field is shown on equally spaced grid points instead of the Chebyshev-Fourier grid so that the regions of strong forcing (very close to the walls) are visible. Both effects have the same pattern of contribution to the driving force.}
  \label{fig:stress_div}
  \end{figure}

The next simplification is based on the observation that the two contributions in Eq.~\eqref{eqn:forcing} act in the same direction as shown in Fig.~\ref{fig:stress_div}. Consequently, only one term with corrected amplitude is retained in Eq.~\ref{eqn:forcing}, which correctly describes the induced flow based on particle density ($\Phi$), polarisation ($\bm{n}$) and the sign of stress intensity ($\sigma_m$). Fig.~\ref{fig:stress_div} further illustrates that the forcing effect is negligible in the bulk region and predominantly influences the flow near the walls. To emphasize this behavior, the figure is presented in a Chebyshev-Fourier space instead of physical space, enabling a clearer visualization of the strong forcing in close proximity to the walls.




\bibliographystyle{plain}
\bibliography{References}
\end{document}